\documentclass[aps,prd,twocolumn,superscriptaddress,showpacs,longbibliography]{revtex4-1}
\usepackage{amsmath}
\usepackage{amssymb}
\usepackage{natbib}
\usepackage{graphicx}
\usepackage{appendix}
\usepackage{hyperref}
\usepackage{url}
\usepackage{longtable}

\begin{document}


\title{Searches for continuous gravitational waves from Scorpius X-1 and XTE J1751-305 
in LIGO's sixth science run}


\author{G.D. Meadors}
\email[]{grant.meadors@ligo.org}
\affiliation{Albert-Einstein-Institut, Max-Planck-Institut f\"ur Gravitations\-physik, D-14476 Potsdam-Golm, Germany}
\affiliation{Albert-Einstein-Institut, Max-Planck-Institut f\"ur Gravi\-ta\-tions\-physik, D-30167 Hannover, Germany }
\author{E. Goetz}
\affiliation{Albert-Einstein-Institut, Max-Planck-Institut f\"ur Gravi\-ta\-tions\-physik, D-30167 Hannover, Germany }
\affiliation{Leibniz Universit\"at Hannover, D-30167 Hannover, Germany }
\affiliation{University of Michigan, Ann Arbor, MI 48109, USA }
\affiliation{LIGO Hanford Observatory, Richland, WA 99352, USA }
\author{K. Riles}
\affiliation{University of Michigan, Ann Arbor, MI 48109, USA }
\author{T. Creighton}
\affiliation{The University of Texas Rio Grande Valley, Brownsville, TX 78520, USA}
\author{F. Robinet}
\affiliation{LAL, Univ. Paris-Sud, CNRS/IN2P3, Universit\'e Paris-Saclay, F-91898 Orsay, France }


\date{\today}

\begin{abstract}
Scorpius X-1 (Sco X-1) and X-ray transient XTE J1751-305 are Low-Mass X-ray Binaries (LMXBs) that may emit continuous gravitational waves detectable in the band of ground-based interferometric observatories.
Neutron stars in LMXBs could reach a torque-balance steady-state equilibrium in which angular momentum addition from infalling matter from the binary companion
is balanced by angular momentum loss, conceivably due to gravitational-wave emission.
Torque-balance predicts a scale for detectable gravitational-wave strain based on observed X-ray flux. 
This paper describes a search for Sco X-1 and XTE J1751-305 in LIGO Science Run 6 data using the TwoSpect algorithm, based on searching 
for orbital modulations in the frequency domain.
While no detections are claimed, upper limits on continuous gravitational-wave emission from Sco X-1 are obtained, 
spanning gravitational wave frequencies from 40 to 2040 Hz and projected semi-major axes from 0.90 to 1.98 light-seconds.
These upper limits are injection-validated, equal any previous set in initial LIGO data, and extend over a broader parameter range.
At optimal strain sensitivity, achieved at 165~Hz, the 95\% confidence level random-polarization upper limit on dimensionless strain $h_0$ is approximately $1.8\times 10^{-24}$.
The closest approach to the torque-balance limit, within a factor of 27, is also at 165 Hz.
Upper limits are set in particular narrow frequency bands of interest for J1751-305.
These are the first upper limits known to date on $r$-mode emission from this XTE source.
The TwoSpect method will be used in upcoming searches of Advanced LIGO and Virgo data.

\end{abstract}

\pacs{04.30.-w, 04.30.Tv, 04.40.Dg, 95.30.Sf., 95.75.Pq, 95.85.Sz, 97.60.Jd}

\maketitle



\section{Introduction\label{introduction}}

Non-axisymmetric spinning neutron stars are predicted to emit continuous gravitational waves (GWs)~\cite{Brady1998}.
In particular, Low-Mass X-ray Binaries (LMXBs) may sustain long-lasting non-axisymmetry through accretion onto the constituent neutron star from the binary partner.
This accretion may reach a torque-balance, where angular momentum carried away by GW emission equals that added from accretion~\cite{PapaloizouPringle1978,Wagoner1984}, depending on the ability of the neutron star equation of state to support non-axisymmetric shapes or fluid modes, as well as the absence of other dissipative mechanisms.
Scorpius X-1 (Sco X-1), the brightest enduring extrasolar X-ray source~\cite{Giacconi1962}, is noteworthy: because detectable GW energy flux scales with observed X-ray flux, it is a promising source for the Laser Interferometer Gravitational-wave Observatory (LIGO) and fellow GW observatories~\cite{Bildsten1998}.
While GW emission due to the $l=m=2$ mass quadrupole moment is expected at twice the neutron star's spin frequency $(f_0 = 2\nu)$, 
emission may occur at other frequencies in the case of $r$-mode \textit{Rossby} oscillations~\cite{Andersson1998,Bildsten1998,Friedman1998,Owen1998}.
Here we carry out a broadband search for continuous GWs from Sco X-1, as well as narrowband searches for $r$-modes
centered on particular frequencies in the case of X-ray transient XTE J1751-305 for which a sharp
spectral line in X-rays has been observed~\cite{Strohmayer2014} that may be indicative of non-radial oscillation modes (such as $r$-modes).

GW detector data have been analyzed using various strategies to search for continuous waves~\cite{Jaranowski1998,HoughTransformKrishnan2004,LSCPulsarS4,LSCPowerFlux2009,PowerFluxMethod2010,PowerFluxAllSky2012}.
The method used here, TwoSpect~\cite{GoetzTwoSpectMethods2011}, has been used opportunistically in a previous all-sky analysis 
to search for Sco X-1 in LIGO Science Run 6 (S6) and the second and third Virgo Science Runs (VSR2 and 3)~\cite{GoetzTwoSpectResults2014}, but it has been subsequently improved (by a factor of $9.5/4.0 = 2.375$ for random polarization~\cite{MeadorsDirectedMethods2016}) and those improvements demonstrated in a Mock Data Challenge (MDC) of simulated signals~\cite{ScoX1MDC2015PRD}. 
This method searches for evidence of orbital modulation of a continuous narrowband signal in strain data as seen in the time-frequency domain.
Upper limits from previous searches for Sco X-1 in initial LIGO data~\cite{AbbottScoX12007,AbadieStoch2011,GoetzTwoSpectResults2014,Sammut2014PRD,Sideband2015} have used diverse techniques to calculate upper limits for each algorithm.
We produce frequentist upper limits using injections in each band, validating against an extrapolated estimate.
Per-band injections yield more conservative limits, which we present, matching the best previous while extending results to a broader band (40 to 2040 Hz).
Sco X-1 ephemerides are listed in Table~\ref{scox1_table_params}.
Data is used from LIGO Science Run 6 (S6), 2009 July 09 to 2010 October 20, with 4-km-long LIGO Hanford (H1) and LIGO Livingston (L1) detectors, as described in Table~\ref{scox1_table_search}.

\begin{table*}[t]
\begin{tabular}{r r r}
Sco X-1 parameter & Value & Units\\
\hline \\
Distance ($d$)~\cite{Bradshaw1999} & $2.8\pm0.3$ & kpc\\
Eccentricity ($\epsilon$)~\cite{ScoX1MDC2015PRD} & $< 0.068$ $(3 \sigma)$ & ---\\
Right ascension ($\alpha$)~\cite{2mass06} & 16:19:55.067 $\pm 0.06'' $ & --- \\
Declination ($\delta$)~\cite{2mass06} & $-15^\circ 38'25.02''\pm 0.06''$ & ---\\
X-ray flux at Earth ($\mathcal{F}_\mathrm{X-ray}$)~\cite{Watts2008} & $3.9\times10^{-7}$ &  erg cm$^{-2}$ s$^{-1}$\\
Orbital period ($P$)~\cite{Galloway2014} & $68023.70 \pm 0.04$ & s\\
Projected semi-major axis ($a_p$)~\cite{2002ApJ...568..273S} & $1.44\pm0.18$ & s\\
\end{tabular}
\caption{
Sco X-1 prior measured parameters from electromagnetic observations (reproduced from methods paper~\cite{MeadorsDirectedMethods2016})
Note that the projected semi-major axis is in units of time, $a_p = (a \sin i)/c$; the value is derived from a velocity amplitude of
$K_1=40\pm5\ {\rm km\,s^{-1}}$ with uncertainty as understood at the time of the search~\cite{AbbottScoX12007,ScoX1MDC2015PRD,Galloway2014}.
Uncertainty has since increased (private communication: derived from the electromagnetically-measured projected radial velocity~\cite{WangSteeghsGalloway2016}).
\label{scox1_table_params}
}
\end{table*}

Because the spin frequency $\nu$~\cite{Markwardt2002,Strohmayer2014} and frequency derivative~\cite{Riggio2011} of XTE J1751-305 are known, its search is significantly less intensive than the search across frequencies required for Sco X-1.
This paper will focus on Sco X-1 to illustrate the general case.

\section{Search Method\label{quant_directed}}
Our search method for Sco X-1 and J1751 is derived from an all-sky algorithm, TwoSpect~\cite{GoetzTwoSpectMethods2011,GoetzTwoSpectResults2014}, specialized to a single sky location~\cite{MeadorsDirectedMethods2016}.

Orbital motion of binary systems involves multiple parameters beyond an isolated continuous GW source.
Here we assume a circular orbit of known
orbital period ($P = 2\pi/\Omega$). 
The algorithm is not sensitive to initial GW phase, nor to orbital
phase as manifested by time of ascension $T_\mathrm{asc}$, the time epoch of the source crossing the orbital ascending node.
Argument of periapsis is also ignored, because orbital eccentricity is small; slight variations could result in biased parameter estimation for signals 
but should not affect upper limits (see Appendix~\ref{eccentricity_section}).
Sky position ($\alpha, \delta$) and period for Sco X-1 are known to sufficient precision~\cite{2mass06,Galloway2014}
to use fixed values in the search~\cite{GoetzTwoSpectMethods2011,MeadorsDirectedMethods2016}.
Spindown $\dot{f_0}$ is presumed small due to torque-balance.
Amplitude parameters (strain $h_0$, neutron star inclination $\iota$, initial GW phase $\Phi_0$, and polarization angle $\psi$) are averaged out through time-dependent antenna functions $F_+$ and $F_\times$.
The antenna functions depend on the observatories~\cite{Jaranowski1998}.

Given known sky location and period, the search is restricted to two dimensions: $f_0$ and projected semi-major axis, $a_p = (a \sin i)/c$.
($a \sin i$ is in units of distance, $a_p$ in units of time).
Doppler frequency modulation depth, $\Delta f_\mathrm{obs} = 2\pi a_p f_0/P$, is more directly observable than $a_p$, and template spacing is uniform in $\Delta f_\mathrm{obs}$, so search grids are specified in ($f_0$, $\Delta f_\mathrm{obs}$).
These parameters describe the assumed phase evolution, $\Phi(t)$, of the strain $h(t)$ for detector time $t$ and solar-system barycenter time $\tau$:

\begin{eqnarray}
h(t) 
&=& \left[F_+ \frac{1+\cos^2 \iota}{2}, F_\times \cos \iota \right]
  \left[ \begin{array}{c} h_0 \cos \Phi(t) \\ h_0 \sin \Phi(t) \end{array}\right],\\
\Phi (t) 
&=& \Phi_0 + 2 \pi f_0 \tau(t) + \Delta f_\mathrm{obs} P \sin \left(\Omega [t - T_\mathrm{asc}] \right),\\
\Delta f_\mathrm{obs} &\equiv& \Omega a_p f_0.
\end{eqnarray}

\noindent A detection statistic, $R$, is constructed based on the strain power, $|h(t)|^2$.
Data is pre-processed into $N$ Short Fourier Transforms (SFTs) of duration $T_\mathrm{coh}$ indexed by $n$ with frequency bins $k$.
As explained previously~\cite{GoetzTwoSpectMethods2011,MeadorsDirectedMethods2016}, the normalized power $P^n_k$ is calculated per SFT, along with its running expectation value, $\left<P_k\right>^n$.
The circular-polarization antenna pattern power, $F_n^2 \equiv F^2_{+,n} + F^2_{\times,n}$ is also computed.
The normalized, background-subtracted, antenna-function-dependent SFT powers $\tilde{P}^n_k$ are calculated for each $(n,k)$ pixel.
Powers $\tilde{P}^n_k$ are then Fourier-transformed by $\mathcal{F}_{f'}$ from the time-domain $n$ to the second-frequency domain, $f'$, which corresponds to orbital period.
Lastly, SFT background noise $\lambda(f')$ is estimated~\cite{GoetzTwoSpectMethods2011}.
This yields second-frequency domain power, $Z$:

\begin{eqnarray}
\tilde{P}^n_k &\equiv& \frac{F_n^2 (P_k^n - \left<P_k\right>^n)}{(\left<P_k\right>^n)^2}\left[\sum\limits_{n'}^N \frac{F_{n'}^4}{(\left<P_k\right>^{n'})^2} \right]^{-1},
\label{equation_with_antenna_pattern}\\
Z_k(f') &=& \frac{\left| \mathcal{F}_{f'} [\tilde{P}^n_k]  \right|^2}{\left< \lambda(f') \right>}.
\label{second_Fourier_power}
\end{eqnarray}

\noindent We reindex $Z_k(f')$ as $Z_i$, where $i$ is a pixel index in the $(k,f')$ plane.
Each pixel has an expected mean $\lambda_i$ and a template weight $w_i$, proportional to the expected magnitude in the presence of a signal.
We sort $i$ in decreasing order of $w_i$ and construct the $R$ statistic from the $M$ highest-weighted pixels:

\begin{equation}
R=\frac{\sum_{i=0}^{M-1}w_{i}[Z_{i}-\lambda_{i}]}{\sum_{i=0}^{M-1}[w_{i}]^{2}}.
\label{TwoSpect_R_statistic}
\end{equation}

\noindent The signal model affects the $R$ statistic through $w_i$, whereas data affects $R$ via $Z_i$.
If one pixel $i$ is dominant, $R$ will approach an exponential distribution; if pixels are equal-magnitude, $R$ will approach a Gaussian normal distribution.
As Equations 13-15 of Goetz~\cite{GoetzTwoSpectMethods2011} show, the reconstructed strain amplitude of a signal $h_\mathrm{rec}$, is proportional to the quarter-root of $R/T_\mathrm{obs}$ for fixed $T_\mathrm{SFT} = T_\mathrm{coh}$.
For the same reason, for a fixed duty cycle and a non-transient signal, we expect $R$ to grow linearly with $T_\mathrm{obs}$.

Previous, all-sky searches using this program~\cite{GoetzTwoSpectResults2014} have not calculated the $R$-statistic for the entire parameter space.
Such a calculation was computationally infeasible, because the parameter space included additional dimensions of period and sky location.
$R$ was used as a follow-up to an initial, \textit{incoherent harmonic sum} stage.
Incoherent harmonic summing involves combining power at each $f'$ with powers at integer multiples of that $f'$: this approximates an optimal search for any signal that varies periodically, but not necessarily sinusoidally, with period $1/f'$.
Because this initial stage's statistic was less sensitive than $R$, an overall gain in sensitivity is expected from bypassing it for a fully-templated $R$-statistic analysis, which is feasible for a known source such as Sco X-1.
A gain of $2.375$ for random polarizations is confirmed by the MDC~\cite{ScoX1MDC2015PRD,MeadorsDirectedMethods2016}.

For Sco X-1, torque-balance predicts, for a 1.4 solar mass, 10 km radius neutron star~\cite{Bildsten1998}, 

\begin{equation}
h_0 \approx 3.5\times 10^{-26} (600~\mathrm{Hz})^{1/2} f_0^{-1/2}.
\end{equation}

\noindent At 50 Hz, $h_0 \approx 1.2\times 10^{-25}$ represents the high end of likely values~\cite{ScoX1MDC2015PRD,MeadorsDirectedMethods2016}. 
Spin-wandering, or fluctuation in $f_0$ due to time-varying accretion rate, is expected, but the analysis coherence time $T_\mathrm{coh}$ is sufficiently short that fluctuations remain within a single Fourier-transform bin and are not likely to affect the search (see Appendix~\ref{spin-wander-bins}).

Using 20\%-mismatch criteria~\cite{GoetzTwoSpectMethods2011} for the $R$-statistic we choose a rectangular-spaced search grid in $f_0$ and $\Delta f_\mathrm{obs}$~\cite{MeadorsDirectedMethods2016}. 
Detector noise and computational cost limit the search to $[f_\mathrm{min}, f_\mathrm{max}] \approx [40, 2040]$~Hz.
SFT coherence times $T_\mathrm{coh}$ are made as long as possible until signals drift out of frequency bins due to Doppler shift from binary orbital motion: 840 s is chosen for $f_0 \in [40, 360]$ Hz and 360 s for $f_0 \in [360, 2040]$ Hz.
Shorter SFTs contain a Doppler-modulated signal in-bin for longer portions of the orbit; this is a particular concern for large values of $a_p$.
Analysis is parallelized into jobs of frequency bandwidth $f_\mathrm{bw} = 0.1$ Hz; $f_\mathrm{bw}$ is the maximum feasible given 2 GB RAM per cluster node.
Each job covers $a_p$ over $\pm 3 \sigma_\mathrm{a_p}$ by stepping through uniform $\Delta f_\mathrm{obs}$.
Equation 6 of the Sco X-1 methods paper~\cite{MeadorsDirectedMethods2016},

\begin{eqnarray}
N_{\mathrm{{template}}} 
 &=& 2 \left(T_\mathrm{coh} + \frac{1}{f_\mathrm{bw}}\right) \nonumber \\
 &\times& \left[ 1+\frac{4 \pi T_\mathrm{coh}}{P} (6\sigma_{a_p})(f_\mathrm{max} + f_\mathrm{min} + f_\mathrm{bw})\right] \nonumber\\
 &\times& (f_\mathrm{max} - f_\mathrm{min}).
\label{N_template_simple}
\end{eqnarray}

\noindent estimates the number of templates required.
Evaluated piecewise for $T_\mathrm{coh} = 840$ s and 360 s over $[40,360]$ and $[360,2040]$ Hz
respectively yields $3.7\times 10^7$ and $2.2\times 10^8$ templates per detector.
Including the separate analyses of H1 and L1 interferometers, the total is approximately $5.1\times10^8$ templates.
Each template returns the $R$-statistic, proportional to $h_0^4$, along with a single-template $\log_{10} p$-value.

Simulations are used to set statistical thresholds compatible with the large number of correlated templates.
Note that the single-template $p$-value ceiling is not corrected for a trials factor appropriate to 2000 Hz. 
This deficiency, which would need resolution in case of future detection, arises from challenges in estimating the effect of long-range correlated structures~\cite{MeadorsDirectedMethods2016}.
Bonferroni correction (multiplying by the number of templates), is excessively conservative.
We defer the issue to a later time, focusing instead on the uncorrected $p$-value corresponding to a particular empirical false alarm rate.
The MDC~\cite{ScoX1MDC2015PRD} found that a detection criterion of single-template $\log_{10} p < -7.75$ present with coincidence in two observatories corresponded to a 5-Hz $p$-value of 0.01. 
(In the MDC, this 5-Hz $p$-value was called a false alarm probability of 1\% per 5 Hz). 
In this search, this detection criterion is interpreted instead as a follow-up criterion.

Methods of setting criteria, parameter estimation in case of detection, and upper limits in its absence, are described in~\cite{MeadorsDirectedMethods2016}.
While these methods suffice for Gaussian noise, real detector data contains artifacts.
Here we detail detection efficiency and validation of upper limits using simulated signals, \textit{injected} into real data.

When $h_0$ upper limits are set in noise power spectral density $S_H$, they can be compared across search algorithms in terms of \textit{sensitivity depth} \cite{LeaciPrixDirectedFStatPRD,BehnkeGalacticCenter2015}:

\begin{equation}
D(f) = S_H^{1/2}(f) h_0^{-1}(f).
\label{sensitivityDepthEq}
\end{equation}

\noindent The sensitivity depth of an algorithm is expected to be roughly constant across varying $S_H(f)$ for fixed $T_\mathrm{coh}$ and with total observing time $T_\mathrm{obs}$.
Search algorithms with higher sensitivity depth than others, given equal $T_\mathrm{obs}$, are said to be more sensitive.

\begin{table*}[t]
\begin{tabular}{l r r r r r}
Search parameter & (H1 840-s SFTs) & (L1 840-s SFTs)  & (H1 360-s SFTs) & (L1 360-s SFTs) & Units\\
\hline \\
S6 start & 931035615 & -- & -- & -- & GPS time (s)\\
S6 end & 971622015 & -- & -- & -- & GPS time (s)\\
Search start & 931052760 & 931052760 & 931071900 & 931071660 & GPS time (s) \\
Search end & 971621820 & 971621880 & 971614500 & 971614680 & GPS time (s) \\
Duration & 40569060 & 40569120 & 40542600 & 40543020 & (s)\\
$f_0$ start & 40.0 & 40.0 & 260.0 & 260.0 & (Hz)\\
$f_0$ end & 360.0 & 360.0 & 2040.0 & 2040.0 & (Hz)\\
Orbital period & 68023.8259 & -- & -- & -- & (s)\\
$a_p$ min & 0.90 & -- & -- & -- & (s)\\
$a_p$ max & 1.98 & -- & -- & -- & (s)
\end{tabular}
\caption{
Parameters for the Sco X-1 search.
Note that different values apply depending on detector (H1, L1) and Short Fourier Transform (SFT) duration $T_\mathrm{coh}=$(360 s, 840 s).
Although 360-s SFTs start from 260 Hz, the [260, 360] Hz band results are reported based on more-sensitive 840-s SFTs. 
Also note that $P = 68023.8259$ s is used in the analysis, based on outdated ephemeris; prior investigations~\cite{GoetzTwoSpectMethods2011} suggest this has negligible effect.
\label{scox1_table_search}
}
\end{table*}

        \subsection{Detection efficiency}
        \label{det_eff_subsection}

\textit{Detection efficiency} is the probability of detecting a signal of a certain strain $h_0$. 
The detector noise floor varies only with $f_0$, so we also marginalize over $a_p$. 
Although the Doppler parameter $a_p$ is a search dimension, it is spanned by at most $\approx 2.9 \times 10^2$ templates, whereas $\approx 1.7 \times 10^6$ templates are required to span $f_0$.
Efficiency is calculated for 0.1 Hz-wide bands of frequency $f_0$, and marginalized over Gaussian-distributed $a_p$ ($\sigma_{a_p} = 0.18$ s) and $P$ ($\sigma_P = 0.0432$ s), as well as uniform-distributed amplitude parameters $(\psi, \Phi_0, \cos \iota)$; sky location $(\alpha, \delta)$ is fixed, and $h_0$ is log-uniform over a factor of 50 range that depends on the estimated noise-floor.

For each 0.1-Hz band, 200 signals are simulated.
Injections are made for each observatory with appropriate antenna pattern and time delay.
A total of $8\times 10^6$ injections are produced (spanning 2000 Hz; 2 detectors).
Injections cover a range of amplitude and Doppler parameters.
These are aggregated into 1-Hz bands for adequate statistics.
Per-injection recovered $R$-statistics at an injection-centered template, and its immediate neighbors, are compared against the loudest $R$-statistic in the 0.1-Hz band without injections.
Centering the injection recovery grid on the actual injection location may result in a slight overestimate of average detection efficiency.
Extrapolation proceeds from an expected mean mismatch $\hat{m} = 1/3$ grid units in any hypercubic lattice~\cite{MessengerLattice2009}.
Each grid unit equals the parameter space distance at which mismatch equals a specified level, $\bar{m}$, for a total mismatch of $m = \bar{m}\hat{m}$.
In~\cite{MessengerLattice2009}, inspired by the $\mathcal{F}$-statistic~\cite{Jaranowski1998}, mismatch is a loss in power, $h_0^2$, but here, it is a loss in power-squared, $h_0^4$.
Brady \textit{et al} Equation 5.4~\cite{Brady1998} connects an offset signal measured by power, $\tilde{h}(\Delta \lambda)$ to a centered signal, $\tilde{h}(0)$: rearranging, $\tilde{h}(\Delta \lambda) = \tilde{h}(0)\sqrt{1-m}$.
($m\propto (\Delta \lambda)^2$ near a maximum).
As the $\mathcal{F}$-statistic is proportional to $\tilde{h} \propto h_0^2$, but for us $R \propto h_0^4$, we square the $|\tilde{h}(\Delta \lambda)|^2/|\tilde{h}(0)|^2$ term in Equation 5.4 to find our mis-estimate in terms of mismatch: $h_0 (\Delta \lambda) = h_0 (0) (1-m)^{1/4}$. 
With our 20\% mismatch giving $\bar{m}=0.2$, the ratio is $(1-0.2/3)^{1/4}\approx 0.983$.
Therefore, we estimate this effect to be approximately~2\%, less than typical calibration uncertainties in previous science runs~\cite{AbadieCalibration2010}.
Injections with $R$ greater than the loudest $R$-statistic are classed as `detections.'

\begin{figure}[t]
\begin{center}
\includegraphics[trim= 0 0 0 15, clip, width=0.40\paperwidth,keepaspectratio]{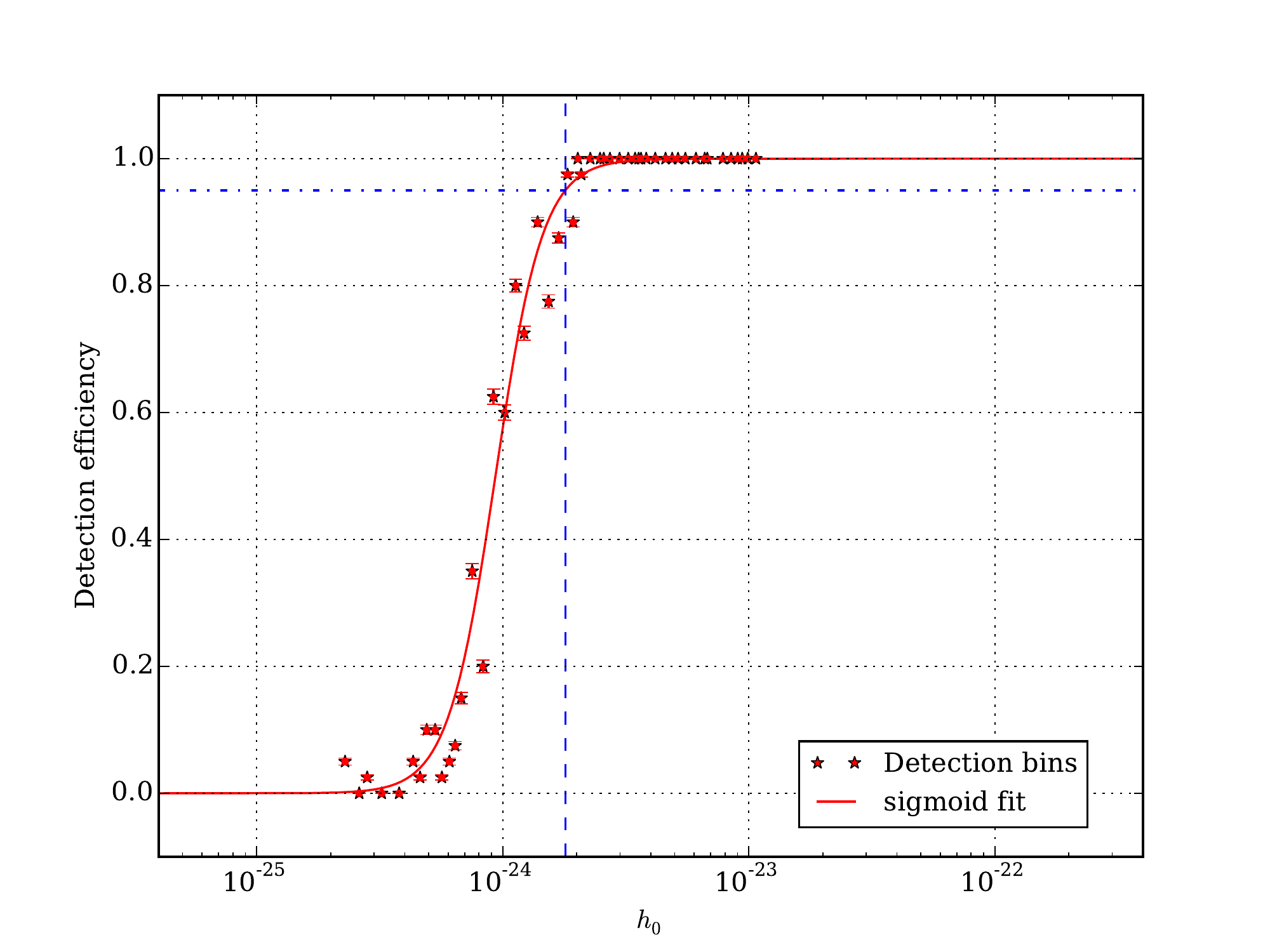}
\caption{
Detection efficiency of 2000 injections into H1 data over $[165.0, 166.0]$ Hz, with varying amplitude and Doppler parameters.
A maximum-likelihood sigmoid fit is made to the unbinned sigmoid distribution of detected/non-detected injections (based on a threshold of $\log_{10} p <$  -7.75); the figure shows binned detection probability estimates for illustration purposes only.
The strain value \textit{(dashed vertical blue line)} yielding 95\% efficiency \textit{(dashed-and-dotted horizontal blue line)} determines the strain upper limit in this band.
}
\label{S6_det_eff_165}
\end{center}
\end{figure}

Detection is expected to become more probable as strain increases, following an approximate sigmoid curve $s(h_0)$.
A two-parameter maximum-likelihood fit is made to $s(h_0)$, from which the 95 \% level is estimated analytically.
Figure~\ref{S6_det_eff_165} plots detection efficiency in the sample band [165.0, 166.0] Hz.

This injection procedure is not identical to the process for identifying detections in real data.
The differences arise from the computational cost of the search and follow-up.
In real data, the detection process begins by comparing against the pre-existing threshold from the MDC~\cite{MeadorsDirectedMethods2016,ScoX1MDC2015PRD}: those with single-template $\log_{10} p < -7.75$ at both detectors are checked for coincidence.
A separate program checks whether $(f_0,\Delta f_\mathrm{obs})$ are within a coincidence requirement of $1/T_\mathrm{coh}$.
Templates that pass are clustered and evaluated in follow-up (Section~\ref{outlier_section}).

        \subsection{Upper limits}
        \label{uls_subsection}

\begin{figure}[t]
\begin{center}
\includegraphics[trim=0 0 0 42, clip, width=0.40\paperwidth,keepaspectratio]{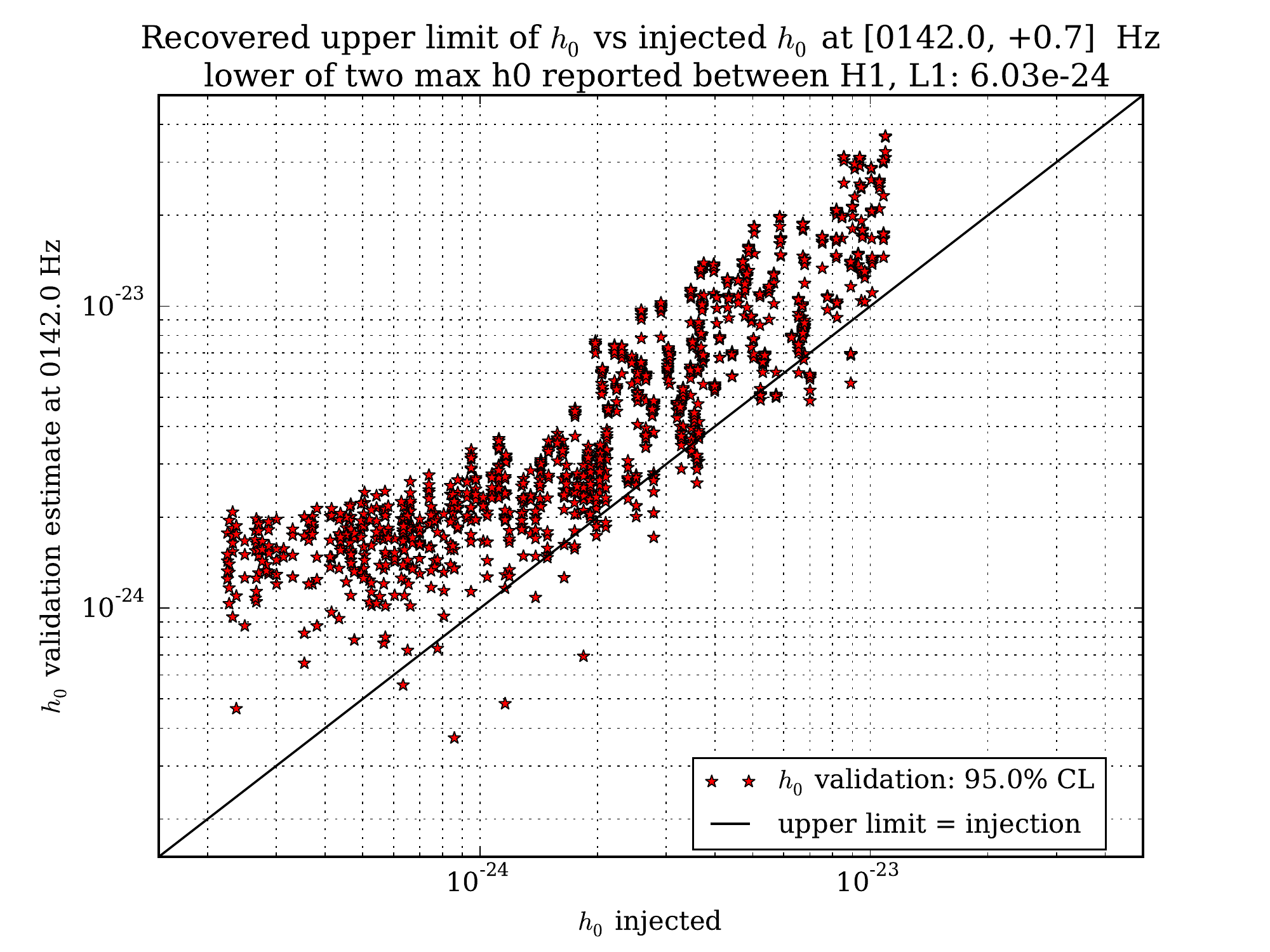}
\caption{
Upper limit validation, estimated $h_0$ vs injected $h_0$: 700 injections into S6 data in the [142.0, 142.7] Hz band.
The uncalibrated `recovered' strain, $h_\mathrm{rec}$, must be calibrated by scaling to an `estimated' $h_0 = \rho_\mathrm{UL} h_\mathrm{rec}$ such that it is greater than or equal to injected $h_0$ at least 95\% of the time\textit{(red points)}.
This must hold for all values of $h_0$.
Applying the scaling factor to the loudest outlier in each band, as described in the text, yields a result consistent with the final 95\% detection efficiency levels, exemplified by Figure~\ref{S6_det_eff_165}, thus validating the upper limits.
}
\label{S6_ULs_142}
\end{center}
\end{figure}

Absent detections, upper limits at the 95\% confidence level  ($h_\mathrm{95\%}$) are the result of the search.
Upper limits are taken as the $95\%$ detection efficiency levels, following prior practice~\cite{FeldmanCousins1998}, computed as in Section~\ref{det_eff_subsection}.
Again, 0.1-Hz bands are aggregated into 1-Hz bands for adequate statistics.
To find $h_\mathrm{95\%}$, we take the sigmoid fits $s_K(h_0)$ for observatory $K$ for a given bin, and analytically invert to obtain $h_0$ for $s_K = 0.95$, and choose the minimum between the two observatories $K$. 
This is then repeated for each 1-Hz bin.

Validation is performed by taking a band (or limited set of bands), estimating in each band the multiplier needed to convert $R$ into $h_0$, applying that multiplier to the loudest template in all other bands, and comparing that product to the $h_\mathrm{95\%}$ found above.
The initial set of bands examined was [142.0, 143.0] Hz, [162.0, 163.0] Hz, and [222.0, 223.0] Hz, with a preliminary set of injections.
To minimize disturbances, [142.0, 142.7] Hz was focused on with the final injection set.
This validation is shown in Figure~\ref{S6_ULs_142}, with an independent set of 700 injections in [142.0, 142.7] Hz.

For validation, amplitude parameters, particularly $\cos \iota$, induce systematic uncertainty into the estimation of $h_0$.
At low values of $h_0$, TwoSpect cannot resolve injections from the noise, and the slope of Figure~\ref{S6_ULs_142} is flat.
At higher values of $h_0$, it recovers injections with $h_\mathrm{rec} \propto R^{1/4}$, linearly proportional to the true value.
Strain estimates, $h_0$,  must be determined using $h_\mathrm{rec}$ with respect to a stated confidence level and injection population.
In the figure, the estimated $h_0$ is plotted.
This $h_0$ is estimated with the smallest coefficient $\rho_\mathrm{UL}$ such that 95\% of $\rho_\mathrm{UL} h_\mathrm{rec}$ for injections, at any value of strain, are greater than or equal to the true $h_0$.
The conversion factor necessary is $\rho_\mathrm{UL} \approx 4.00$.
Moreover, $\rho_\mathrm{UL}$ can be factored as $2.3\times \rho_{\cos \iota}$, where $\rho_{\cos \iota} = 1.74 (\pm 0.37)$ is a population-dependent estimate of the average ratio of true $h_0$ to recovered $h_0$ given a uniform distribution of $\cos \iota$~\cite{ScoX1MDC2015PRD,MeadorsDirectedMethods2016}. 

The results of multiplying the loudest template $h_\mathrm{rec}$ in all other bands by $\rho_\mathrm{UL}$ are generally consistent with the $h_{95\%}$ found by the detection-efficiency method.
Variation is expected, since the former is an extrapolation from the $\rho_\mathrm{UL}$ estimated for [142.0, 142.7] Hz being uniformly applied across all other bands.
Such variation is permissible when the detection-efficiency method is more conservative.
Agreement can be quantified by comparing the median $\rho_\mathrm{UL} h_\mathrm{rec,i}$ of 10 validation bins, 0.1 Hz each, to the corresponding 1-Hz detection-efficiency bin, $h_\mathrm{95\%,i}$.
Where $r$ is the ratio $\rho_\mathrm{UL} h_\mathrm{rec,i}/h_\mathrm{95\%,i}$, the median $r = 0.755$, mean $0.753$ with standard deviation $0.153$.
At the most sensitive frequency, $r = 0.898$, whereas the 40-50 Hz mean $r=0.851$ and 2030-2040 Hz mean $r=0.692$.
The tendency for smaller $r$ at high frequency may stem from the higher trials factor as $f$ increases, requiring a search over a larger $\Delta f_\mathrm{obs}$ space.
As $h_\mathrm{95\%}$ is larger and more conservative on average, the upper limits are validated.

Upper limits can be set for most frequency bands.
A few bands are consistently too noisy, as identified by statistical tests, and therefore cannot be analyzed.
SFTs must pass a Kolmogorov-Smirnov and Kuiper's test:
non-Gaussian or anomalously noisy data are not used.
These tests are detailed in the TwoSpect all-sky observational paper~\cite{GoetzTwoSpectResults2014}.
For the Sco X-1 search, the 60 Hz and first three harmonic lines at (120, 180, 240) Hz, as well as frequencies near the violin modes around 340 to 350 Hz, are thus excluded~\cite{S6DetChar2015}.
The 40 to 360 Hz H1 and L1 searches excluded cumulative bandwidths of 16.4 Hz and 16.2 Hz respectively, while the 360 to 2040 Hz H1 search excluded 21.4 Hz and L1 16.9 Hz.
For generating upper limits, only 288 bins (0.1 Hz each) could not be analyzed with data from either interferometer.
Aggregating into 1-Hz bins allows setting upper limits even near many of the disturbances, in some sense recovering the bandwidth at the cost of coarser results, by having more injections for better sigmoid fit statistics.
Only 5 aggregated bins of 1 Hz each could not be determined from either interferometer.
Results are detailed in Section~\ref{random_pol_uls}.

\section{Sco X-1 results\label{directed_results}}

\begin{figure}[t]
\begin{center}
\includegraphics[trim=0 0 0 0, clip, width=0.40\paperwidth,keepaspectratio]{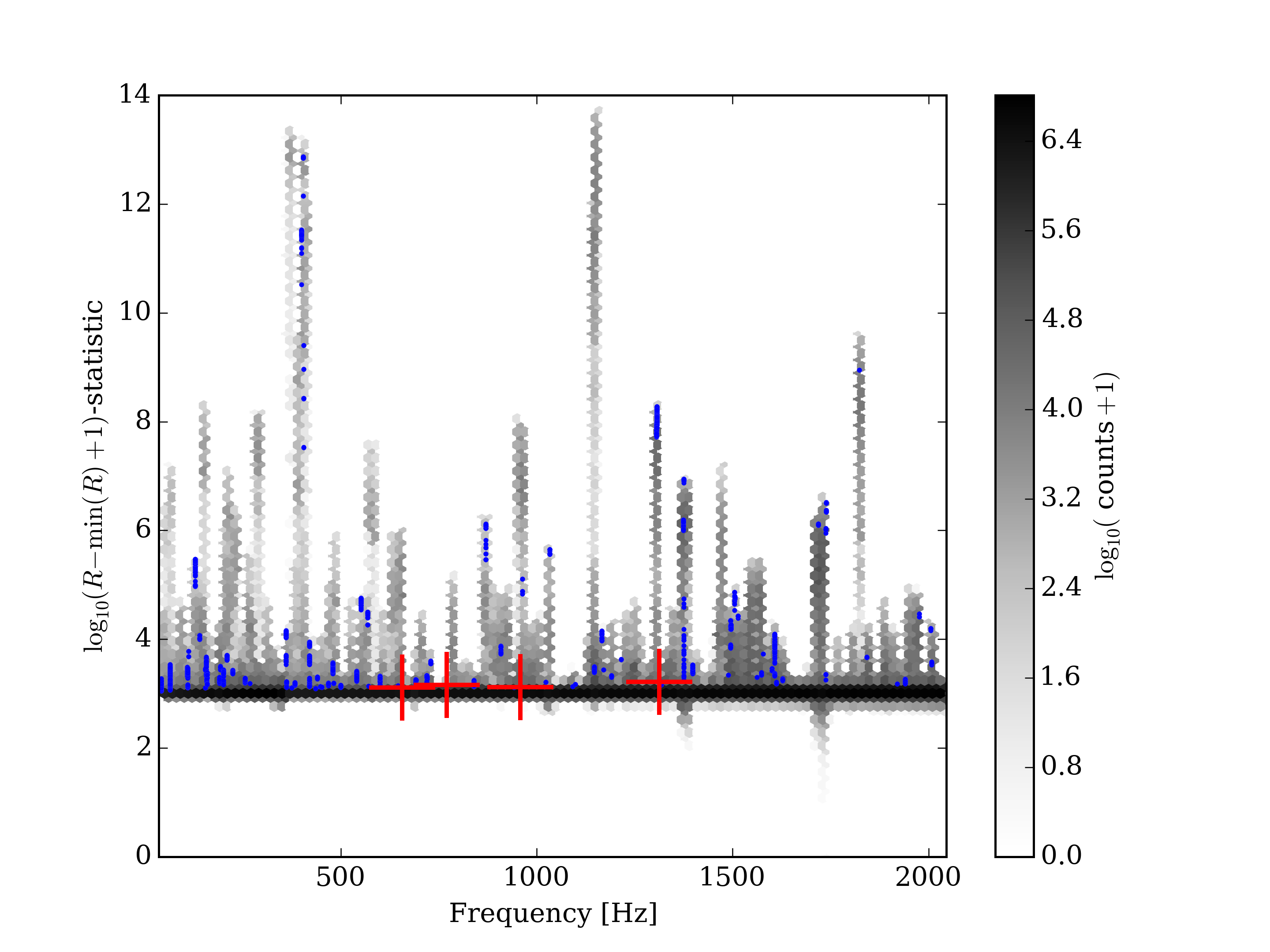}
\caption{
2D histogram with hexagonal bins of logarithmic $R$-statistic versus frequency $f_0$ for the Sco X-1 S6 search.
Histogram for all templates \textit{(gray hex bins)} and followed-up coincident templates only \textit{(blue dots)}.
Variance in $R$ increases with $f_0$, because more pixels are incorporated into the statistic.
However, $R$ remains zero-mean.
Line artifacts are present at many frequencies, extending to $R \approx 5\times 10^{13}$.
The four outliers from Table~\ref{ScoX1S6outlierTable} are marked \textit{(red crosshairs)}; they are at 656, 770, 957, and 1312 Hz.
}
\label{rPlotFig}
\end{center}
\end{figure}

\begin{figure}[t]
\begin{center}
\includegraphics[trim=0 0 0 0, clip, width=0.40\paperwidth,keepaspectratio]{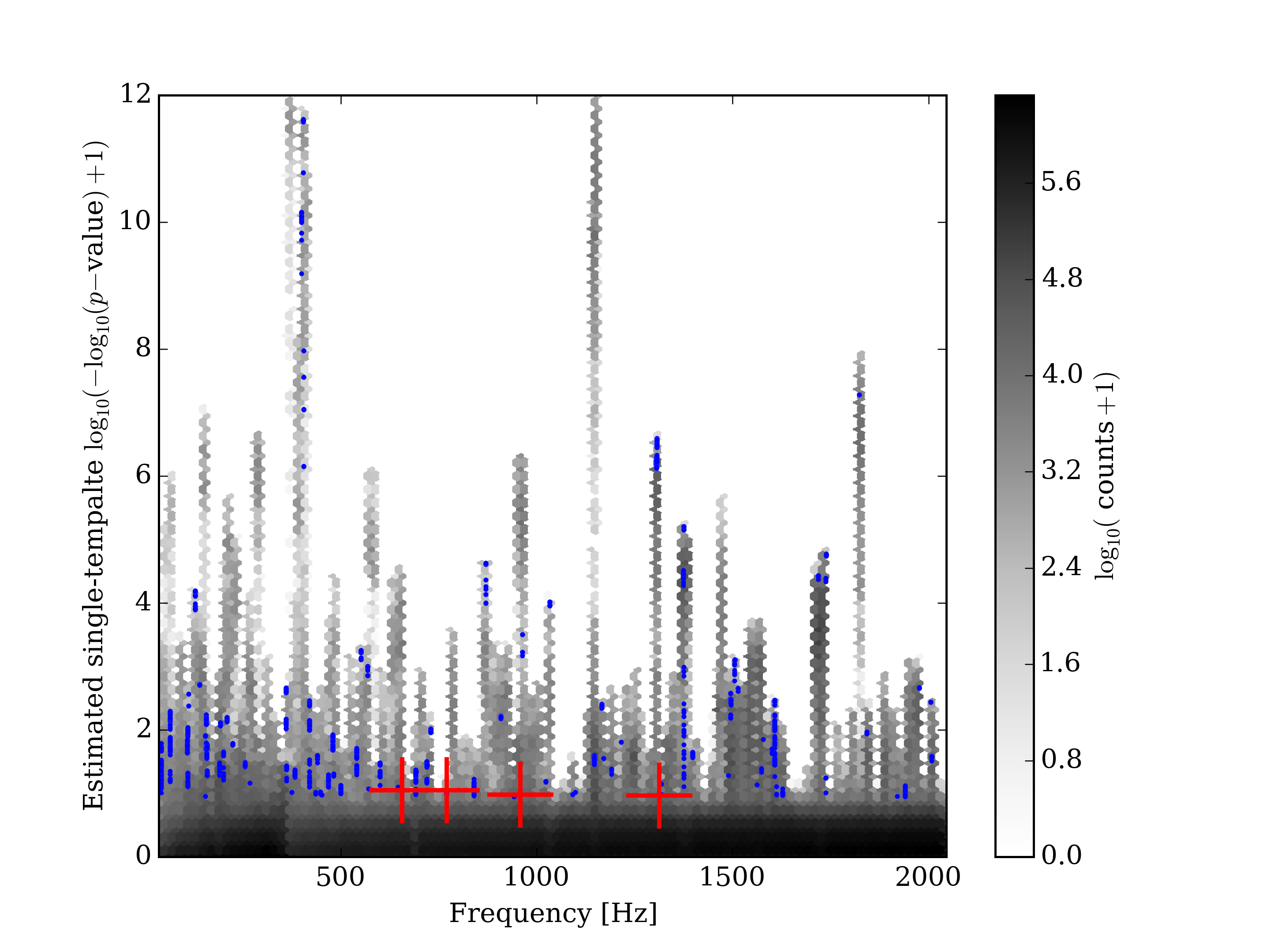}
\caption{
2D histogram with hexagonal bins of doubly-logarithmic (single-template) $p$-value versus frequency $f_0$ for the Sco X-1 S6 search.
Histogram for all templates \textit{(gray hex bins)} and followed-up coincident templates only \textit{(blue dots)}.
Line artifacts align with those in Figure~\ref{rPlotFig}.
The four outliers from Table~\ref{ScoX1S6outlierTable} are marked \textit{(red crosshairs)}; they are at 656, 770, 957, and 1312 Hz.
}
\label{fPlotFig}
\end{center}
\end{figure}

\begin{figure}[t]
\begin{center}
\includegraphics[trim=0 0 0 0, clip, width=0.40\paperwidth,keepaspectratio]{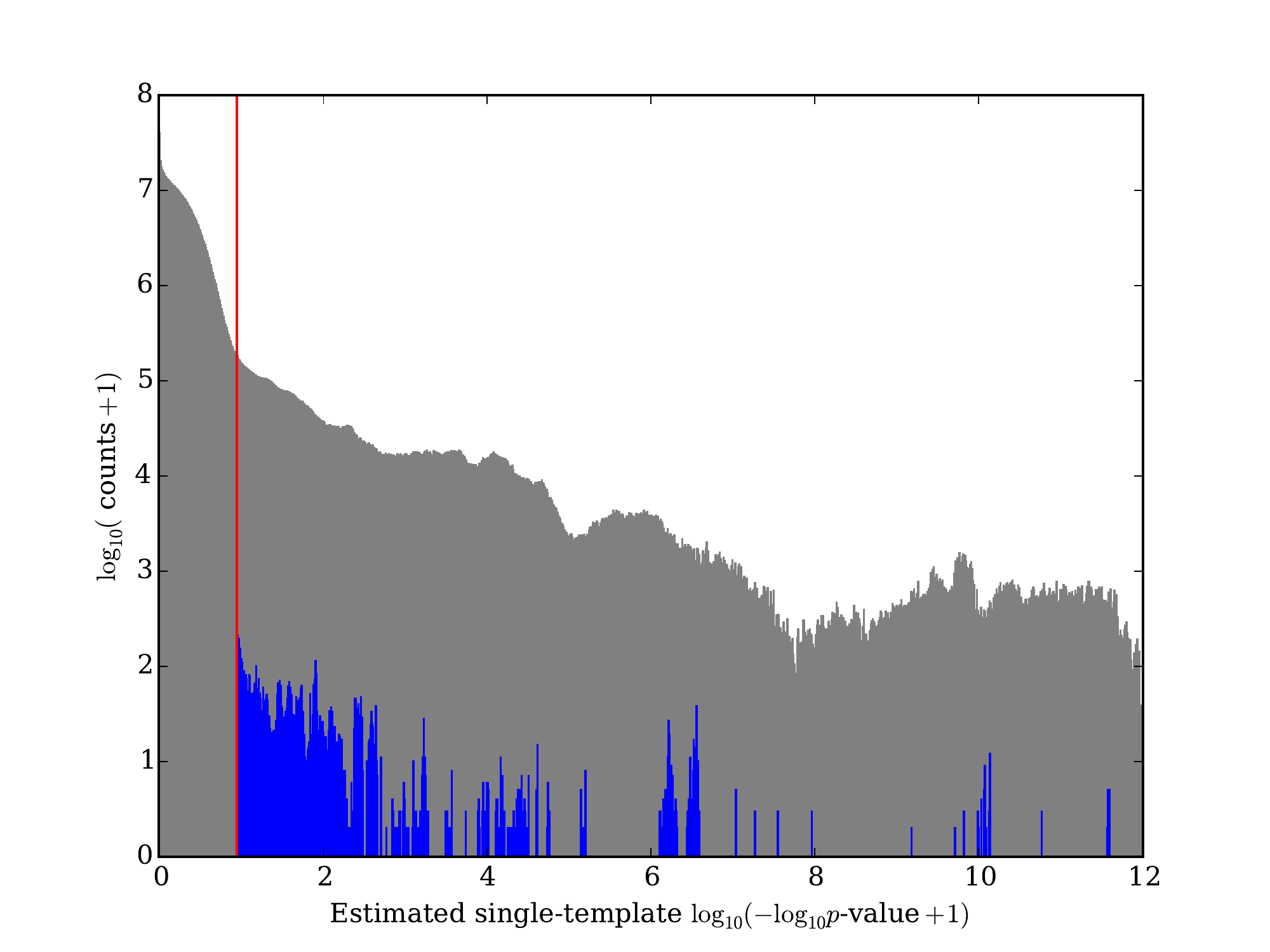}
\caption{
Doubly-logarithmic horizontal scale for the distribution of $-\log_{10}(p)$-values.
Histogram for all templates \textit{(gray)} and followed-up coincident templates only \textit{(blue)}.
For high values, $-\log_{10}(p) \propto R$.
Note that these extreme values of $R$ are, as shown in Figure~\ref{fPlotFig}, typically related to line artifacts in the data.
Follow-up threshold of $\log_{10}p = -7.75$ shown in \textit{(red)}: a template in each observatory must reach this threshold and and be coincident with the other observatory to qualify for follow-up.
In comparison to theoretical expectations, templates with many equally-weighted pixels would have Gaussian-distributed $R$, while templates dominated by a single pixel will have exponentially-distributed $R$.
On this figure's axes, both such distributions would appear as concave-downward curves.
The knee in the slope and extended right tail imply that extreme $-\log_{10}(p)$-values are part of distinct, unmodeled populations, such as the aforementioned line artifacts.
In the absence of artifacts or signals, the below-threshold slope would continue.
}
\label{pLogHistFog}
\end{center}
\end{figure}

Summary results for the $R$-statistic and estimated single-template $p$-value of the Sco X-1 search can be found in Figure~\ref{rPlotFig} through~\ref{pLogHistFog}.
These histograms of the data show structural features for both the entire set of templates as well as those passing threshold and coincidence requirements.

\subsection{Sco X-1 outliers\label{outlier_section}}

Templates matching the statistical threshold in both observatories and coincident within $1/T_\mathrm{coh}$ in both $f_0$ and $\Delta f_\mathrm{obs}$ are clustered together.
Because $f_0$ is a much larger dimension in our search than $\Delta f_\mathrm{obs}$, only $f_0$ is used to define clusters: any points within twice the maximum possible modulation depth (to allow for degeneracies in the parameter space: points on the same $a_p$ vs $f$ structure~\cite{MeadorsDirectedMethods2016}) plus 5 SFT bins (for safety with FFT signal leakage~\cite{GoetzTwoSpectMethods2011}) are considered a cluster.
Of the 90 clustered outliers in Appendix~\ref{SectionAllOutlierTable}, Table~\ref{AllOutlierTable}, 86 are dismissed by visual inspection of the amplitude spectral density in the band.
Most show identifiable artifacts, such as instrumental lines and power harmonics.
Table~\ref{ScoX1S6outlierTable} presents four outliers, present in both interferometers between 40 and 2040 Hz, that do not overlap identifiable artifacts.

\begin{table*}[t]
\begin{tabular}{r r c c c c l}
H1 $f_0$ (Hz) & L1 $f_0$ (Hz) & H1 $\Delta f_\mathrm{obs}$ (Hz) & L1 $\Delta f_\mathrm{obs}$ (Hz) & H1 $h_0$& L1 $h_0$ & Comment \\
\hline
656.6431 & 656.6458 & 0.0650 & 0.0630 & $ 1.48\times 10^{-24}$ & $1.86\times 10^{-24} $ & dismissed by coherent sum\\
770.2250 & 770.2264 & 0.1229 & 0.1256 & $ 2.04\times 10^{-24} $ & $2.46 \times 10^{-24} $ & dismissed by coherent sum \\
957.6972 & 957.6958 & 0.0803 & 0.0817 & $ 2.24\times 10^{-24} $ & $2.88 \times 10^{-24} $& dismissed by coherent sum \\
1312.4542 & 1312.4528 & 0.2373 & 0.2380 & $3.77\times 10^{-24} $ & $4.53 \times 10^{-24} $ & fluctuation (see below) \\
\end{tabular}
\caption{Estimated parameters of outliers not corresponding to obvious artifacts.
Frequency $f_0$, modulation depth $\Delta f_\mathrm{obs}$, and naive recovered $h_0$ are shown.
86 of 90 outliers (clustered templates matching threshold in both H1 and L1) can be easily dismissed due to artifacts or visible disturbances in the amplitude spectral density.
These 4 remaining outliers survive.
Three are dismissed by failing the test of having a higher statistic in when SFTs are coherently summed.
The last is highly unphysical, not self-consistent, and statistically marginal, as described in the text.
Full outlier listing in Table~\ref{AllOutlierTable}.
  }
\label{ScoX1S6outlierTable}
\end{table*}

Of the surviving four outliers in Table~\ref{ScoX1S6outlierTable}, all except the last (outlier 66, 1312.453 Hz) can be dismissed by coherently summing SFTs before calculating the $R$-statistic.
In this case, when H1 and L1 are in simultaneous operation, SFTs from both are phase-shifted to account for detector separation, the matched SFTs are added together, and then analysis proceeds as for a single (virtual) detector.
Real signals are expected to yield higher $R$-statistics using the coherent sum.
This results in higher sensitivity: an H1-L1 sum with unknown $\cos \iota$ and $\psi$ should improve by approximately 29 percent~\cite{TwoSpectCoherentGoetz2015CQG} over single-detector analyses, in the all-sky search.
Directed searches, such as Sco X-1, are not fully characterized, nor are the false alarm and false dismissal probabilities of the test for higher joint-$R$, but the example of the $\mathcal{F}$-statistic multi-detector statistic~\cite{CutlerMulti2005} is informative.
The single-detector $\mathcal{F}$-statistic~\cite{Jaranowski1998} has an expected statistic, $\mathcal{F}$, proportional to a non-centrality parameter $\rho \propto h^2$; with $N$ combined detectors, $\rho \propto N$, so sensitivity to $h$ scales like $\sqrt{N}$. 
Einstein@Home, for example, vetoes candidates for which any single-detector statistic is less than the joint-detector statistic (the $\mathcal{F}$-statistic consistency veto)~\cite{EinsteinHomeS52013}.
Because $R$ is not coherent, it will scale more slowly than $\mathcal{F}$, but should grow with additional detectors.
Only the last outlier, 1312.453 Hz, does have a larger $R$ with coherent-summing.

Multiple considerations suggest that the 1312.453 Hz outlier is nevertheless not a real signal from Sco X-1.
First, note that the follow-up criterion, as noted in Section~\ref{quant_directed}, yield a false alarm probability of 1\% per 5 Hz band in Gaussian noise~\cite{ScoX1MDC2015PRD}.
The data set contains 400 bands of 5 Hz, implying a $(1- 0.99^{400})=0.98$ probability of at least one false alarm.
Given the high false alarm probability of the search's follow-up criterion, it is less likely that any particular outlier arises from actual GW emission from Sco X-1 (presumed to be monochromatic and with no confusion from GW backgrounds). 
Moreover, the $R$-statistic of this outlier does not grow linearly in time if the observation is subdivided.
Indeed, subdividing into quarters or thirds yields inconsistent results for the time interval with the loudest $R$, compared to dividing into halves.
These results are also inconsistent with the expectations that $R \propto T_\mathrm{obs}$ and the corollary that detectable $h_0$ scales with $T_\mathrm{obs}^{-1/4}$ (see Equation~\ref{TwoSpect_R_statistic} and below).
It is possible, however, that the marginal nature of the outlier elevates the false dismissal probability of this test.

Alternately, Sco X-1 might not be in torque-balance.
Then previous assumptions might not apply, and a transient signal for part of the run could not be ruled out. 
The recovered strain $h_\mathrm{rec}$ is at least 159 times larger than torque-balance: $2.37\times 10^{-26}$ according to prior formulae based on the X-ray flux~\cite{Bildsten1998,MeadorsDirectedMethods2016}. 
(The true strain $h_0$, after correcting for $\cos \iota$, would be even larger on average).
Hence, spin-down would occur at an estimated rate of at least $3.2\times 10^{-7}$ Hz/s (4\% of a frequency bin per coherence time; 4674 bins during the observing time).
This rapid frequency drift should induce either an extended cluster of outliers at various frequencies consistent with evolution during the run, or else might fail to produce an outlier at all, if the accumulated power in each bin is insufficient.
Both possibilities are inconsistent with observed results, so the signal is presumably unphysical.

Although the possibility remains open for a signal at $h_0$ greater than torque-balance predictions, it would require more statistically consistent evidence to substantiate -- the false alarm probability of the search and the inconsistent behavior of the $R$-statistic do not provide this evidence.
Statistical fluctuation is the most likely explanation for the 1312.453 Hz outlier: we conclude that no signals from Sco X-1 have been detected.

\subsection{Sco X-1 random-polarization upper limits}
\label{random_pol_uls}

\begin{figure*}[t]
\begin{center}
\includegraphics[trim=0 0 0 0, clip, width=0.70\paperwidth,keepaspectratio]{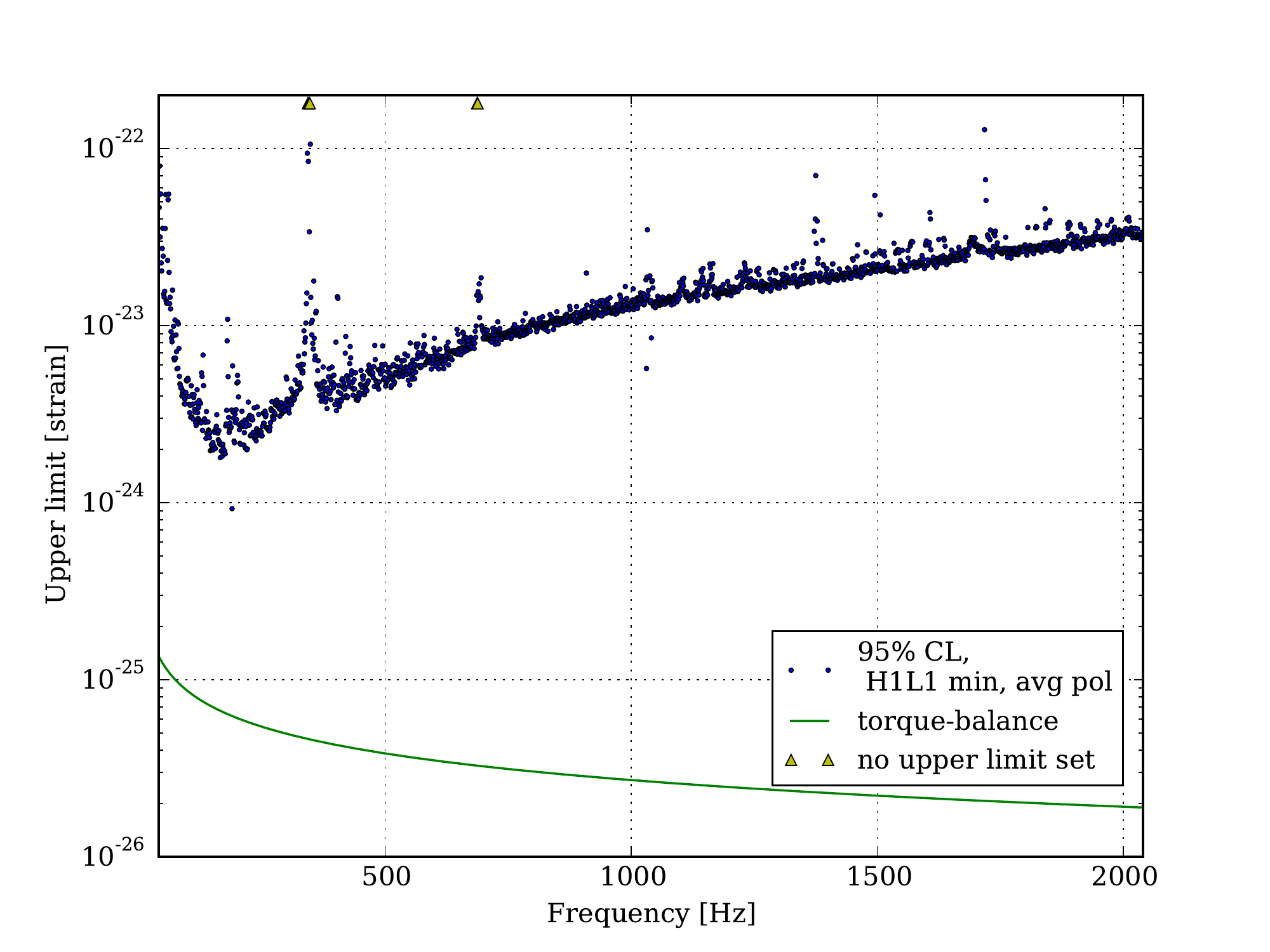}
\caption{
Upper limit for average (random) polarization GW from Sco X-1 in S6, joint H1-L1, at 95\% confidence $\textit{(blue dots)}$.
Validated by determining the 95\% detection efficiency in injections, this spectrum covers $[40, 2040]$ Hz, using the lower upper limit from either observatory when both yielded results.
Results are for 1-Hz bands (closed lower edge, open upper edge).
Three bands (189, 1031, and 1041 Hz) have anomalously low upper-limit that probably stem from spectral artifacts visible the run-averaged SFT amplitude spectral density.
Five near-overlapping bands total are marked \textit{(yellow triangles)} where upper limits could not be set:
four (343, 347, 687 and 688) Hz could not be fit because of additional artifacts yielding insufficient data for the search method, and 345 Hz could not be numerically determined by maximum likelihood.
The region around 345 Hz contains the first harmonic of the interferometer suspension violin mode, responsible for these disturbances.
Bifurcation in high-frequency limits arises from certain bands being contaminated in H1 and limit being set by the less-sensitive L1 interferometer.
Results make no assumptions on $\cos \iota$.
Note: torque-balance \textit{(green line)} assumes a 1.4 solar mass, 10 km neutron star.
}
\label{S6_H1L1_UL}
\end{center}
\end{figure*}

Upper limits (ULs) for the [40, 2040] Hz spectrum, with 95\% confidence given random polarization, in units of dimensionless strain, are shown in Figure~\ref{S6_H1L1_UL}, produced as explained in Section~\ref{uls_subsection}.
The minimum value, $1.8 \times 10^{-24}$, is achieved at 165 Hz: it is approximately 27 times the torque-balance strain limit at that frequency.

Previous Radiometer searches using LIGO Science Run S5 data~\cite{AbadieStoch2011} must be converted for comparison.
Note that although S6 noise was improved over S5, the observation time $T_\mathrm{obs}$ of S5 was approximately twice as long as S6.
Whereas both searches should improve proportional to the quarter root of $T_\mathrm{obs}$, these effects approximately cancel.
The Radiometer UL is calculated for 90\% confidence given circular (optimal) rather than random polarization; it is presented in units of $h_\mathrm{RMS}$, which do not directly correspond to physical strain.
Signal leakage across the 0.25 Hz Radiometer bins affects $h_\mathrm{RMS}$ near the boundaries of bins for signals below 538 Hz (and all signals above)~\cite{ScoX1MDC2015PRD}.
The effects of signal leakage and polarization dependence have been investigated and found to scale $h_\mathrm{RMS} \approx 0.48 h_0$, on average, over the range of [50, 1500]~Hz~\cite{ScoX1MDC2015PRD}.
For a specific $df=0.25$-Hz bin, the conversion is $h_0 = C_\mathrm{cp} \times \sqrt{C_\mathrm{mbf} \times \hat{Y}_\mathrm{tot}df}$, where $C_\mathrm{cp} = 1.74\pm0.37$ and $C_\mathrm{mbf}$ (frequency-dependent) are new with respect to S5. 
Applying factors of 1.74 and $C_\mathrm{mbf}(160\mathrm{ Hz}) = (1.2)^{1/2}$ to the best claimed Radiometer result, $h_\mathrm{RMS} = 7\times 10^{-25}$ near 160 Hz, implies the 90\%-confidence S5 upper limit was $h_0^{90\%}\approx 1.3 \times 10^{-24}$.
Compare the average factor, $(0.48)^{-1} \times 7\times 10^{-25} = 1.5\times 10^{-24}$, or the Sideband group's estimated conversion factor~\cite{Sideband2015}, $(2.43)\times 7\times 10^{-25} = 1.7\times 10^{-24}$.
We can estimate our 90\%-confidence levels for direct comparison.
The 165 Hz detection efficiency sigmoid (Figure~\ref{S6_det_eff_165}) that yields our best upper limits likewise reaches 0.90 at approximately $1.5\times 10^{-24}$ in a 1-Hz bin group.
Our 0.1-Hz injection sets do not facilitate 0.25-Hz Radiometer bins, but grouping by 0.3-Hz for sigmoid fits yields 90\%-confidence limits of $(1.3\times 10^{-24}, 1.7\times 10^{-24}, 1.6 \times 10^{-24})$ for bands respectively starting $(165.0, 165.3, 165.6)$ Hz.
We believe our limits to be at least comparable to the Radiometer S5 limits, but spanning a broader frequency range and more rigorously calculated, using Monte Carlo injections for the full range of astrophysical polarizations, presented in physical units of strain $h_0$ at 95\% confidence.


Sideband searches in S5 data~\cite{Sideband2015} are more analogous.
However, the S5 UL is produced using a Bayesian method different from the MDC and from our frequentist Monte Carlo injections; some discrepancy may thus arise.
The Sideband UL over the range [50, 550] Hz is calculated at 95\% confidence, including both random polarization, as here, as well as a restricted prior.
Its semicoherent (sensitivity $\propto T_\mathrm{obs}^{1/2}$) analysis spans 10 days.
When comparing these results, note that the Sideband paper uses the median UL from within a 1 Hz band.
Our method instead generates a single UL for each 1 Hz band.
As this means that each fit in the current paper may be adversely affected by noise within the band, the quantities are not directly equivalent.
For example, calculating 95\%-sigmoid ULs from our 0.1-Hz bins and taking the median for $[165.0, 165.9]$-Hz yields a median of $1.66\times 10^{-24}$ compared to our stated UL of $1.8\times 10^{-24}$ on the sigmoid for the whole 1 Hz.
The UL at our most sensitive frequency, 165 Hz, is nevertheless between the median ($1.3\times 10^{-24}$) and worst Sideband UL at its most sensitive frequency, 150 Hz.
Moreover, the results in the current paper cover a broader range of [40, 2040] Hz.

While method differences still complicate comparison, we find our ULs to be within factor of 1.4 between Radiometer S5 and Sideband S5 limits; while not distinctly better in sensitivity, they span a larger parameter space (2 kHz in $f_0$) in S6 data. 

\section{XTE J1751-305}

\subsection{XTE J1751-305 method}

Discovered in 2002~\cite{Markwardt2002}, the X-ray transient J1751-305 is another binary system with potential for continuous GW emission.
In 2014~\cite{Strohmayer2014}, X-ray observations of J1751 were reported that exhibited signs consistent with non-radial oscillation modes, such as $r$- (and \textit{gravity} $g$-) modes.
Debate ensued; an $r$-mode might have already spun J1751 down below detectable levels~\cite{Andersson2014}, but a crust-only surface $r$-mode might not and could still be present~\cite{Lee2014}. 
J1751 is thus an interesting candidate for an opportunistic search well-suited to our method.

J1751 has the shortest known orbital period of any X-ray binary: $P \approx 2545.3$ s.
Its semi-major axis is $a_p$ $\approx0.010$ s. 
Crucially, its spin frequency is known: $\nu = 435.31799$ Hz.
J1751 is distant, at an estimated $d > 7$ kpc, near the galactic center.
We consider GW emission at the spin frequency $f_0 = \nu$, the relativistic-corrected $r$-mode frequency $f_0 = (2-0.5727597)\nu$~\cite{Strohmayer2014} (dependent on the unknown equation of state), and the quadrupolar frequency $f_0 = 2\nu$.
Bands of $f_\mathrm{band} = 2.0$ Hz, centered approximately on each frequency, and $a_p$ $\pm 0.0033$ s, are selected with a single period $P = 2545.3414$ s.
The search is run on 200-s and 240-s SFTs to cope with high Doppler acceleration from the short $P$.
This analysis overcovers uncertainties, yet is practical ($<10^5$ templates) for a single-processor in less than a day.
Parameters are listed in Table~\ref{xte_table_params}.

\begin{table}[t]
\begin{tabular}{r r r}
XTE J1751-305 parameter & Value & Units\\
\hline \\
Distance ($d$)~\cite{Markwardt2002} & $>7$ & kpc\\
Eccentricity ($\epsilon$)~\cite{Markwardt2002} & $< 1.7 \times 10^{-3}$ & ---\\
Right ascension ($\alpha$)~\cite{Markwardt2002} & 17:51:13.49 $\pm 0.05'' $ & --- \\
Declination ($\delta$)~\cite{Markwardt2002} & $-30^\circ 37'23.4''\pm 0.6''$ & ---\\
Orbital period ($P$)~\cite{Markwardt2002} & $2545.3414 \pm 0.0038$ & s\\
Proj. semi-major axis ($a_p$)~\cite{Markwardt2002} & $ 10.1134 \pm 0.0083$ & ms\\
Frequency derivative $\dot{\nu}_\mathrm{spin}$~\cite{Riggio2011} & $-0.55(12) \times 10^{-14}$  & Hz s$^{-1}$\\
Spin frequency $\nu_\mathrm{spin}$~\cite{Markwardt2002} & 435.317993681(12) & Hz\\
$r$-mode $f_0$~\cite{Strohmayer2014} &  621.3034 & Hz\\
$2\nu_\mathrm{spin}$ & 870.63598 & Hz
\end{tabular}
\caption{
Orbital and spin parameters of XTE J1751-305.
\label{xte_table_params}
}
\end{table}

\begin{table}[t]
\begin{tabular}{r r}
Frequency (Hz) & Strain (dimensionless) \\
\hline \\
$435.3$ & $3.2656 \times 10^{-24}$\\
$621.3$ & $4.7125 \times 10^{-24}$\\
$870.6$ & $7.7532 \times 10^{-24}$
\end{tabular}
\caption{
Upper limits for 0.1 Hz frequency bands containing three possible putative signals from J1751-305.
In increasing order of frequency, these are the $\nu_\mathrm{spin}$, $r$-mode, and $2\nu_\mathrm{spin}$.
The upper limits represent an average (random) polarization 95\%-confidence limit, determined using 200 injections per observatory for each band.
\label{xte_table_results}
}
\end{table}

\subsection{XTE J1751-305 results}

No evidence for GW emission from XTE J1751-305 is found in the frequency bands [434.5, 436.5] Hz, [620.5, 622.5] Hz, or [869.5, 871.5] Hz.
No candidate templates passed both threshold and coincidence requirements between H1 and L1 observatories.

In the absence of any candidates, upper limits are set in Table~\ref{xte_table_results}.
We follow the same method as Section~\ref{random_pol_uls}, with minor modifications: sky location, period, and projected semi-major axis are adjusted to fit J1751, and the frequency span is limited to the 0.1 Hz bands containing each of three possible emission frequencies: $\nu_\mathrm{spin}$, $r$-mode, and $2\nu_\mathrm{spin}$.
To our knowledge, these are the first limits set on J1751 using gravitational-wave data.

Debate~\cite{Lee2014,Andersson2014} over the existence of $r$-modes in J1751 motivates future searches.
S6 amplitude spectral densities ranged around $[3-6]\times 10^{-23}$ Hz$^{-1/2}$ over [435, 870] Hz.
In the same range, Advanced LIGO in O1 has achieved strain sensitivities of $[1-2] \times 10^{-23}$ Hz$^{-1/2}$~\cite{GW150914LIGO}; design sensitivity may achieve $[4-5]\times10^{-24}$ Hz$^{-1/2}$~\cite{HarryALIGO2010}. 
Extrapolating to equivalent duration design sensitivity data, we might anticipate approximately order-of-magnitude improvements in upper limits.
Andersson \textit{et al}~\cite{Andersson2014} discuss how, with a nominal internal mode amplitude $\alpha = 10^{-3}$, the expected strain of an $r$-mode in LIGO might be approximately $1\times 10^{-24}$.
While internal mode amplitude is also highly-uncertain, future upper limits could indeed be illuminating.

\section{Conclusions}

LIGO S6 data is analyzed for continuous gravitational wave emission from Sco X-1 using the TwoSpect method. While no credible detections are made, upper limits are set for randomly polarized gravitational waves from Sco X-1 from 40 to 2040~Hz.
This analysis covers the uncertainty in projected semi-major axis, $\sigma_{a_p}$, over $\pm 3 \sigma_{a_p}$ as was known in S6~\cite{2002ApJ...568..273S}, though $\sigma_{a_p}$ ephemerides are evolving~\cite{WangSteeghsGalloway2016}.
Upper limits for randomly polarized GWs are set for this 2-kHz frequency band, except for five 1-Hz disturbed bands.
The best upper limit is $h_0 = 1.8\times 10^{-24}$ at 165 Hz, 27 times the torque-balance limit there.
XTE J1751-305 is also targeted and shows no sign of GW emission, although transients cannot be ruled out; we set the first upper limits on J1751 using GW data.

These results using TwoSpect~\cite{GoetzTwoSpectMethods2011,MeadorsDirectedMethods2016} are the best that span a 2 kHz frequency range and use initial LIGO data~\cite{AbbottScoX12007,AbadieStoch2011,GoetzTwoSpectResults2014,Sammut2014PRD,Sideband2015}.
In this paper, the frequency range surveyed is considerably larger than the [20,~57.25] Hz previously-analyzed by TwoSpect for S6 and Virgo VSR2/3~\cite{GoetzTwoSpectResults2014} or [50, 550] Hz analyzed by Sideband for S5~\cite{Sideband2015}.
Moreover, in an advance over Radiometer S5 limits~\cite{AbadieStoch2011}, the limits presented here are in physical units of strain $h_0$ (instead of $h_\mathrm{RMS}$), for simulated random polarizations, at the 95\%-confidence level.
The relative performance of the pipelines is fairly consistent with that observed in the Sco X-1 Mock Data Challenge~\cite{ScoX1MDC2015PRD}, allowing for differences in bin size and observation time for Radiometer and the Bayesian upper limit technique used for Sideband in S5.
Already, as this paper shows, TwoSpect achieves a sensitivity depth of approximately $10$ Hz$^{-1/2}$ (as previously estimated~\cite{MeadorsDirectedMethods2016}; see Equation~\ref{sensitivityDepthEq}) with respect to the S6 amplitude spectral density.
Extrapolated to an equivalent duration of Advanced LIGO design sensitivity data at an amplitude spectral density of $4 \times 10^{-24}$ Hz$^{-1/2}$, this method might set an upper limit of $4\times 10^{-25}$ on Sco X-1, well within an order of magnitude of the predicted torque-balance limit~\cite{Bildsten1998}.
Promising pipelines are in development~\cite{Dhurandhar2008,ScoX1CrossCorr2015PRD}, though further enhancement to the TwoSpect algorithm is also expected~\cite{TwoSpectCoherentGoetz2015CQG}.
This method sets new upper limits on Sco X-1 GW emission using S6 data, and it is ready to be applied to data from Advanced LIGO, Advanced Virgo, and KAGRA.

\begin{acknowledgments}
This work was partly funded by National Science Foundation grants NSF PHY-1505932 and NSF HRD-1242090 as well as by the Max-Planck-Institut. 
These investigations use data and computing resources from the LIGO Scientific Collaboration.
Further thanks to the Albert-Einstein-Institut Hannover and the Leibniz Universit\"{a}t Hannover for support of the Atlas cluster on which most of the computing for this project was done.
The data used in this analysis is available through the LIGO Open Science Center~\cite{S6LOSC}.
Thanks to Maria Alessandra Papa, Reinhard Prix, and Chris Messenger for proofreading and suggestions, as well as to our referees for thorough reading and comments.
This document bears LIGO Document Number DCC-P1500039.
\end{acknowledgments}

\appendix

\section{Eccentricity\label{eccentricity_section}}
Eccentricity effects have not been analytically considered for this algorithm before.
Starting from the instantaneous relative GW frequency offset $(\delta f)/f$~\cite{LeaciPrixDirectedFStatPRD},
with eccentric anomaly $E$, eccentricity $e$ and argument of periapse $\omega$,

\begin{equation}
\left|\frac{\delta f}{f}\right|= a_\mathrm{p} \Omega \left|\frac{\sqrt{1-e^2} \cos E \cos \omega - \sin E \sin \omega}{1-e \cos E} \right|.
\label{irfo}
\end{equation}

\noindent
Instantaneous frequency offsets result in shifted estimates of the apparent intrinsic frequency and modulation depth: labeling these offsets $\delta f_a$ and $\delta \Delta f_a$,

\begin{eqnarray}
\delta f_\mathrm{a} &=& \frac{1}{2} \left| \max(\delta f) + \min(\delta f)\right|,
\label{dfa}\\
\delta \Delta f_\mathrm{a} &=& \frac{1}{2} \left| \max(\delta f) - \min(\delta f)\right| - \Delta f_\mathrm{obs}.
\label{ddfa}
\end{eqnarray}

\noindent Extremizing over $E$ to solve where $e \geq 0$, $e \ll 1$,

\begin{eqnarray}
\frac{\delta f_\mathrm{a}}{\Delta f_\mathrm{obs}} &=& \left( \frac{\sqrt{1-e^2} \cos^2 \omega + \sin^2 \omega}{1-e^2 \cos^2 \omega} e \cos \omega \right),
\label{solved_dfa}\\
\frac{\delta \Delta f_\mathrm{a}}{\Delta f_\mathrm{obs}} &=& \left( \frac{\sqrt{1-e^2} \cos^2 \omega + \sin^2 \omega}{1-e^2 \cos^2 \omega} \right) - 1.
\label{solved_ddfa}
\end{eqnarray}

\noindent The offsets are related by $\delta f_a = \left(\delta \Delta f_a + \Delta f_\mathrm{obs}\right) e \cos \omega$. 
Equations~\ref{solved_dfa} and~\ref{solved_ddfa} predict the $(f, \Delta f_\mathrm{obs})$ position of $\max(R)$.
A case is simulated at the edge of the Sco X-1 search parameter space, with $a_p = 1.95$ light-s, $f = 2039.95$ Hz; compared to when $e=0$, the $e=0.07$ simulation shows maximum shifts of $\delta f_a = -25.7$ mHz, $\delta \Delta f_a = +0.9$ mHz for $\omega=0$ and $\delta f_a = +25.7$ mHz, $\delta \Delta f_a = +0.9$ mHz for $\omega = \pi$.
This is consistent with predictions up to a sign, which depends on the convention for $\omega$.

The $R$ value computed for an injected signal with a simulated eccentricity  $\sigma_e$ is comparable to that when $e=0$, so given the physical probability of orbital circularization, eccentricity will be treated as a problem in parameter estimation but not while setting upper limits. 

\section{Spin-wandering across frequency bins\label{spin-wander-bins}}

Consider a gravitational-wave signal that changes in frequency $f_0$ by one frequency bin, width $1/T_\mathrm{coh}$, during the observing time, $T_\mathrm{obs}$.
Model this change as linear in time; for example, consider an LMXB where accretion suddenly stopped, and the neutron star kept radiating GW at its torque-balance frequency $f_0$.
This model should set an upper limit.
The limit is found by comparing the frequency bin width allowance to the spindown rate.
The frequency bin width allowance is,

\begin{eqnarray}
\frac{d f_0}{dt} T_\mathrm{obs} &=& -\frac{1}{T_\mathrm{coh}}.
\end{eqnarray}

Comparing to the spindown rate of an isolated star~\cite{Riles2013},

\begin{eqnarray}
\frac{dE}{dt} &=& -\frac{32}{5} \frac{G}{c^5} I_{zz}^2 \epsilon^2 (\pi f_0)^6,\\
h_0 &=& 4\pi^2 \frac{G I_{zz} f_0^2 \epsilon}{c^4 r}.
\end{eqnarray}

\noindent The first equation can be transformed by noting that $E = I_{zz} (\pi f_0)^2/2$, so

\begin{equation}
\frac{d f_0}{dt} = \frac{1}{\pi^2 I_{zz} f_0},
\end{equation}

\noindent and we can readily solve for $\epsilon$ in terms of $h_0$.
Combining these solutions,

\begin{equation}
\frac{d f_0}{dt} = -\frac{2 c^3}{5 G I_{zz}} r^2 h_0^2 f_0.
\end{equation}

Using values of $I_{zz} = 10^{38}$ $\text{ kg~m}^2$ and $r = 2.7$ $\text{kpc}$,

\begin{equation}
\frac{d f_0}{dt} = \left[-1.1 \times 10^{-9} \text{ Hz s}^{-1}\right] \left(\frac{h_0}{10^{-24}} \right)^2\left(\frac{f_0}{100\text{ Hz}} \right).
\end{equation}

For a loud candidate, for example $h_0 = 4\times 10^{-24}$ at $1312.45$ Hz, this spindown rate is $2.4\times 10^{-7}\text{ Hz s}^{-1}$, equal to $95$ Hz spindown over the 40.5 million second $T_\mathrm{obs}$ of S6. This is, respectively, 80 thousand or 34 thousand times larger than the width of, respectively, 840-s or 360-s SFTs. Moreover, for a signal well above the torque-balance limit, the linear spindown approximation might be valid.

Applying the same logic to a signal at the torque-balance limit,

\begin{eqnarray}
\frac{d f_0}{dt} &=&  -1.1 \times 10^{-9} \text{ Hz s}^{-1} \left[\frac{ 3.5\times 10^{-26}}{10^{-24}} \right]^2 \frac{600}{100},\\
 &=& -8.2 \times 10^{-12} \text{ Hz s}^{-1},
\end{eqnarray}

\noindent or approximately 0.033 Hz spindown over S6.
This is also wider than the width of either frequency bin, by a factor of approximately 28 for 840-s SFTs and 12 for 360-s SFTs.
However, this is a worst-case scenario, and torque-balance is thought to be an equilibrium condition.
The sudden spindown of a neutron star from torque-balance is not expected.
Because the potential for spin-wandering effects is not completely negligible, however, this subject remains a topic of active investigation~\cite{MukherjeeSpinWandering2016}.

\section{Full list of outliers}
\label{SectionAllOutlierTable}
Please see Table~\ref{AllOutlierTable} for a full list of outlier clusters.

\newpage

\begin{longtable*}{r|r|r|r|r|r|r|r|r|r|r|r}
\# & H1 $f$ (Hz) & $\Delta f_\mathrm{obs}$ (Hz) & $R$ & $h_\mathrm{rec}$ & $-\log_{10}p$ & L1 $f$ (Hz) & $\Delta f_\mathrm{obs}$ (Hz) & $R$ & $h_\mathrm{rec}$ & $-\log_{10}p$ & Follow-up\\[0.1ex]
\hline\\[0.1ex]
\endfirsthead
\multicolumn{12}{@{}l}{\tablename~\thetable~--~\textit{continued from previous page}}\\[0.1ex]
\hline\\[0.1ex]
\# & H1 $f$ (Hz) & $\Delta f_\mathrm{obs}$ (Hz) & $R$ & $h_\mathrm{rec}$ & $-\log_{10}p$ & L1 $f$ (Hz) & $\Delta f_\mathrm{obs}$ (Hz) & $R$ & $h_\mathrm{rec}$ & $-\log_{10}p$ & Follow-up\\[0.1ex]
\hline\\[0.1ex]
\endhead
\hline\\[0.1ex]
\multicolumn{12}{r@{}}{\tablename~\thetable~--~\textit{continued on next page}}\\[0.1ex]
\endfoot
\endlastfoot
    1 &   42.0125 &  0.0077 &  8.42e+02 &  6.60e-23 &  -5.98e+01 &    42.0113 &  0.0077 &  1.82e+02 &  1.52e-23 &  -1.59e+01 &   a\\[0.1ex]
    2 &   63.9946 &  0.0114 &  2.35e+03 &  6.49e-24 &  -1.94e+02 &    63.9946 &  0.0114 &  2.95e+02 &  3.66e-24 &  -2.18e+01 &   b\\[0.1ex]
    3 &  108.1006 &  0.0195 &  2.01e+03 &  2.46e-24 &  -6.84e+01 &   108.1012 &  0.0189 &  2.15e+02 &  9.08e-25 &  -9.81e+00 &   a\\[0.1ex]
    4 &  108.8673 &  0.0184 &  1.26e+03 &  1.55e-24 &  -7.42e+01 &   108.8685 &  0.0196 &  1.89e+03 &  1.55e-24 &  -1.07e+02 &   c\\[0.1ex]
    5 &  109.4732 &  0.0183 &  4.26e+02 &  1.05e-24 &  -1.97e+01 &   109.4726 &  0.0174 &  2.43e+02 &  9.29e-25 &  -1.07e+01 &   a\\[0.1ex]
    6 &  111.0202 &  0.0128 &  4.97e+03 &  2.07e-24 &  -3.63e+02 &   111.0214 &  0.0116 &  1.42e+02 &  8.09e-25 &  -7.75e+00 &   b\\[0.1ex]
    7 &  128.0149 &  0.0234 &  2.96e+05 &  4.31e-24 &  -1.55e+04 &   128.0149 &  0.0231 &  2.97e+02 &  1.12e-24 &  -1.41e+01 &   b\\[0.1ex]
    8 &  139.5190 &  0.0166 &  2.52e+02 &  6.93e-25 &  -1.34e+01 &   139.5179 &  0.0178 &  8.75e+03 &  1.72e-24 &  -5.14e+02 &   b\\[0.1ex]
    9 &  154.0411 &  0.0267 &  1.70e+03 &  9.73e-25 &  -8.01e+01 &   154.0423 &  0.0255 &  2.37e+02 &  6.64e-25 &  -9.09e+00 &   b\\[0.1ex]
   10 &  154.5571 &  0.0256 &  2.30e+02 &  5.89e-25 &  -7.96e+00 &   154.5560 &  0.0268 &  2.30e+02 &  6.61e-25 &  -7.82e+00 &   a\\[0.1ex]
   11 &  156.8167 &  0.0237 &  3.67e+02 &  6.71e-25 &  -1.51e+01 &   156.8179 &  0.0225 &  3.35e+03 &  1.27e-24 &  -1.70e+02 &   a\\[0.1ex]
   12 &  157.9946 &  0.0268 &  8.22e+02 &  8.10e-25 &  -3.71e+01 &   157.9935 &  0.0256 &  4.89e+02 &  7.79e-25 &  -2.13e+01 &   b\\[0.1ex]
   13 &  158.3673 &  0.0269 &  4.56e+02 &  6.98e-25 &  -1.98e+01 &   158.3685 &  0.0278 &  3.02e+02 &  6.91e-25 &  -1.09e+01 &   b\\[0.1ex]
   14 &  158.8619 &  0.0276 &  1.27e+03 &  8.93e-25 &  -5.72e+01 &   158.8619 &  0.0285 &  6.50e+02 &  8.37e-25 &  -2.71e+01 &   b\\[0.1ex]
   15 &  190.8000 &  0.0322 &  7.45e+02 &  9.48e-25 &  -2.39e+01 &   190.8012 &  0.0334 &  4.22e+02 &  8.97e-25 &  -1.34e+01 &   a\\[0.1ex]
   16 &  192.5482 &  0.0184 &  2.11e+03 &  1.12e-24 &  -1.28e+02 &   192.5470 &  0.0172 &  4.91e+02 &  8.13e-25 &  -2.79e+01 &   c\\[0.1ex]
   17 &  200.5298 &  0.0331 &  1.41e+03 &  1.42e-24 &  -4.21e+01 &   200.5310 &  0.0343 &  4.77e+02 &  1.01e-24 &  -1.35e+01 &   a\\[0.1ex]
   18 &  209.2839 &  0.0326 &  3.73e+02 &  8.42e-25 &  -1.30e+01 &   209.2827 &  0.0338 &  3.92e+03 &  1.33e-24 &  -1.54e+02 &   a\\[0.1ex]
   19 &  223.6625 &  0.0373 &  1.63e+03 &  9.31e-25 &  -6.02e+01 &   223.6637 &  0.0385 &  6.58e+02 &  1.28e-24 &  -1.77e+01 &   b\\[0.1ex]
   20 &  256.0327 &  0.0448 &  8.55e+02 &  8.63e-25 &  -2.82e+01 &   256.0327 &  0.0448 &  5.04e+02 &  8.48e-25 &  -1.44e+01 &   b\\[0.1ex]
   21 &  268.1399 &  0.0383 &  2.93e+02 &  6.57e-25 &  -7.99e+00 &   268.1405 &  0.0392 &  4.87e+02 &  1.03e-24 &  -1.34e+01 &   a\\[0.1ex]
   22 &  360.0028 &  0.0638 &  9.95e+03 &  3.94e-24 &  -4.00e+02 &   360.0014 &  0.0652 &  1.34e+04 &  8.41e-24 &  -4.58e+02 &   b\\[0.1ex]
   23 &  361.4792 &  0.0536 &  6.14e+02 &  1.58e-24 &  -2.60e+01 &   361.4792 &  0.0536 &  2.45e+02 &  1.29e-24 &  -1.01e+01 &   a\\[0.1ex]
   24 &  375.3653 &  0.0624 &  2.48e+02 &  1.17e-24 &  -9.33e+00 &   375.3653 &  0.0596 &  2.30e+02 &  1.33e-24 &  -8.03e+00 &   a\\[0.1ex]
   25 &  383.1903 &  0.0701 &  5.58e+02 &  1.28e-24 &  -2.10e+01 &   383.1917 &  0.0694 &  5.05e+02 &  2.27e-24 &  -1.82e+01 &   b\\[0.1ex]
   26 &  400.1000 &  0.0641 &  2.96e+11 &  1.88e-22 &  -1.42e+10 &   400.1000 &  0.0641 &  3.25e+02 &  1.71e-24 &  -1.14e+01 &   b\\[0.1ex]
   27 &  403.7778 &  0.0426 &  6.43e+04 &  4.05e-24 &  -3.73e+03 &   403.7792 &  0.0405 &  6.99e+12 &  5.20e-22 &  -4.12e+11 &   b\\[0.1ex]
   28 &  404.8000 &  0.0720 &  3.50e+04 &  3.46e-24 &  -1.55e+03 &   404.8000 &  0.0740 &  2.51e+09 &  7.33e-23 &  -9.36e+07 &   b\\[0.1ex]
   29 &  420.0111 &  0.0691 &  7.62e+03 &  3.74e-24 &  -2.76e+02 &   420.0097 &  0.0705 &  2.54e+03 &  2.60e-24 &  -9.06e+01 &   b\\[0.1ex]
   30 &  435.2556 &  0.0465 &  2.09e+02 &  1.06e-24 &  -9.24e+00 &   435.2569 &  0.0486 &  1.97e+02 &  1.23e-24 &  -8.26e+00 &   a\\[0.1ex]
   31 &  440.1000 &  0.0805 &  9.89e+02 &  1.55e-24 &  -3.83e+01 &   440.1014 &  0.0791 &  8.92e+02 &  1.85e-24 &  -3.45e+01 &   b\\[0.1ex]
   32 &  448.0861 &  0.0771 &  2.94e+02 &  1.14e-24 &  -9.14e+00 &   448.0875 &  0.0792 &  3.02e+02 &  1.56e-24 &  -9.45e+00 &   b\\[0.1ex]
   33 &  450.9569 &  0.0735 &  2.60e+02 &  1.18e-24 &  -8.39e+00 &   450.9542 &  0.0762 &  2.74e+02 &  1.48e-24 &  -8.16e+00 &   a\\[0.1ex]
   34 &  468.1000 &  0.0745 &  2.99e+02 &  1.40e-24 &  -9.77e+00 &   468.1000 &  0.0745 &  5.20e+02 &  1.60e-24 &  -1.82e+01 &   b\\[0.1ex]
   35 &  480.0167 &  0.0829 &  1.11e+03 &  1.98e-24 &  -3.68e+01 &   480.0194 &  0.0802 &  2.59e+03 &  3.26e-24 &  -8.27e+01 &   b\\[0.1ex]
   36 &  482.2069 &  0.0673 &  2.65e+02 &  1.23e-24 &  -8.90e+00 &   482.2042 &  0.0680 &  5.04e+02 &  1.83e-24 &  -1.91e+01 &   a\\[0.1ex]
   37 &  500.0722 &  0.0915 &  3.82e+02 &  1.41e-24 &  -1.09e+01 &   500.0722 &  0.0915 &  3.82e+02 &  1.69e-24 &  -1.12e+01 &   a\\[0.1ex]
   38 &  539.9583 &  0.0856 &  1.56e+03 &  2.35e-24 &  -5.01e+01 &   539.9556 &  0.0884 &  3.47e+02 &  1.71e-24 &  -9.64e+00 &   b\\[0.1ex]
   39 &  552.0417 &  0.1003 &  3.20e+02 &  1.45e-24 &  -8.37e+00 &   552.0417 &  0.0982 &  5.37e+04 &  5.97e-24 &  -1.77e+03 &   b\\[0.1ex]
   40 &  568.1000 &  0.0935 &  3.05e+04 &  4.44e-24 &  -9.99e+02 &   568.1000 &  0.0915 &  3.42e+02 &  1.81e-24 &  -9.86e+00 &   b\\[0.1ex]
   41 &  570.3514 &  0.0759 &  3.36e+02 &  1.47e-24 &  -1.07e+01 &   570.3528 &  0.0779 &  2.51e+02 &  1.62e-24 &  -7.99e+00 &   a\\[0.1ex]
   42 &  600.0042 &  0.0840 &  3.61e+02 &  1.67e-24 &  -1.03e+01 &   600.0042 &  0.0819 &  9.09e+02 &  2.95e-24 &  -2.93e+01 &   b\\[0.1ex]
   43 &  646.5264 &  0.0822 &  2.71e+02 &  1.57e-24 &  -8.02e+00 &   646.5264 &  0.0850 &  3.68e+02 &  2.02e-24 &  -1.14e+01 &   a\\[0.1ex]
   44 &  656.6431 &  0.0650 &  2.20e+02 &  1.48e-24 &  -7.85e+00 &   656.6458 &  0.0630 &  2.67e+02 &  1.86e-24 &  -1.01e+01 &   d\\[0.1ex]
   45 &  691.1500 &  0.0777 &  4.25e+02 &  2.87e-24 &  -1.47e+01 &   691.1514 &  0.0763 &  6.12e+02 &  3.21e-24 &  -1.99e+01 &   b\\[0.1ex]
   46 &  692.1653 &  0.1133 &  4.62e+02 &  2.50e-24 &  -1.19e+01 &   692.1653 &  0.1154 &  7.05e+02 &  3.13e-24 &  -2.06e+01 &   b\\[0.1ex]
   47 &  719.9819 &  0.0815 &  4.39e+02 &  2.19e-24 &  -1.33e+01 &   719.9806 &  0.0836 &  8.26e+02 &  4.59e-24 &  -2.65e+01 &   b\\[0.1ex]
   48 &  729.6000 &  0.0918 &  2.74e+03 &  3.11e-24 &  -1.02e+02 &   729.5986 &  0.0912 &  4.97e+02 &  2.55e-24 &  -1.54e+01 &   b\\[0.1ex]
   49 &  770.2250 &  0.1229 &  4.18e+02 &  2.04e-24 &  -1.02e+01 &   770.2264 &  0.1256 &  3.98e+02 &  2.46e-24 &  -8.51e+00 &   d\\[0.1ex]
   50 &  839.9417 &  0.1515 &  4.52e+02 &  2.46e-24 &  -8.13e+00 &   839.9403 &  0.1488 &  7.14e+02 &  3.36e-24 &  -1.57e+01 &   b\\[0.1ex]
   51 &  870.0042 &  0.0925 &  4.66e+02 &  2.46e-24 &  -1.41e+01 &   870.0069 &  0.0938 &  1.22e+06 &  2.15e-23 &  -4.20e+04 &   b\\[0.1ex]
   52 &  908.9139 &  0.1330 &  4.54e+02 &  2.52e-24 &  -9.78e+00 &   908.9125 &  0.1358 &  6.19e+03 &  6.16e-24 &  -1.62e+02 &   a\\[0.1ex]
   53 &  942.7431 &  0.0812 &  2.69e+02 &  2.87e-24 &  -7.77e+00 &   942.7458 &  0.0791 &  2.54e+02 &  2.75e-24 &  -7.80e+00 &   a\\[0.1ex]
   54 &  957.6972 &  0.0803 &  2.61e+02 &  2.24e-24 &  -8.04e+00 &   957.6958 &  0.0817 &  2.79e+02 &  2.88e-24 &  -8.47e+00 &   d\\[0.1ex]
   55 &  963.2042 &  0.1615 &  4.52e+02 &  2.58e-24 &  -8.11e+00 &   963.2069 &  0.1595 &  1.27e+05 &  1.36e-23 &  -3.17e+03 &   b\\[0.1ex]
   56 & 1022.9708 &  0.1232 &  5.92e+02 &  4.25e-24 &  -1.44e+01 &  1022.9681 &  0.1260 &  4.13e+02 &  3.31e-24 &  -9.01e+00 &   a\\[0.1ex]
   57 & 1033.9875 &  0.1870 &  4.40e+05 &  3.06e-23 &  -1.04e+04 &  1033.9861 &  0.1891 &  1.46e+03 &  4.77e-24 &  -3.02e+01 &   b\\[0.1ex]
   58 & 1091.4764 &  0.0928 &  3.03e+02 &  2.69e-24 &  -8.58e+00 &  1091.4750 &  0.0942 &  3.10e+02 &  3.40e-24 &  -8.48e+00 &   a\\[0.1ex]
   59 & 1098.2125 &  0.1343 &  4.30e+02 &  3.20e-24 &  -9.39e+00 &  1098.2139 &  0.1357 &  3.88e+02 &  3.72e-24 &  -7.82e+00 &   b\\[0.1ex]
   60 & 1147.6944 &  0.2051 &  1.51e+03 &  4.66e-24 &  -2.80e+01 &  1147.6944 &  0.2023 &  1.98e+03 &  5.76e-24 &  -3.76e+01 &   b\\[0.1ex]
   61 & 1166.1347 &  0.2098 &  4.29e+03 &  6.30e-24 &  -9.00e+01 &  1166.1347 &  0.2119 &  1.29e+04 &  9.09e-24 &  -2.51e+02 &   b\\[0.1ex]
   62 & 1171.0847 &  0.1996 &  4.91e+02 &  3.36e-24 &  -7.76e+00 &  1171.0833 &  0.1975 &  1.72e+03 &  5.38e-24 &  -3.46e+01 &   a\\[0.1ex]
   63 & 1190.6125 &  0.1740 &  4.59e+02 &  3.16e-24 &  -8.20e+00 &  1190.6111 &  0.1719 &  1.10e+03 &  4.88e-24 &  -2.33e+01 &   b\\[0.1ex]
   64 & 1216.0986 &  0.2176 &  5.74e+02 &  3.57e-24 &  -8.42e+00 &  1216.0972 &  0.2196 &  3.18e+03 &  6.72e-24 &  -6.27e+01 &   b\\[0.1ex]
   65 & 1306.6861 &  0.1454 &  4.95e+02 &  3.76e-24 &  -1.05e+01 &  1306.6847 &  0.1426 &  1.51e+08 &  1.02e-22 &  -3.90e+06 &   b\\[0.1ex]
   66 & 1312.4542 &  0.2373 &  5.86e+02 &  3.77e-24 &  -8.17e+00 &  1312.4528 &  0.2380 &  6.07e+02 &  4.53e-24 &  -8.08e+00 &   d\\[0.1ex]
   67 & 1318.7181 &  0.1625 &  4.24e+02 &  3.72e-24 &  -7.78e+00 &  1318.7194 &  0.1611 &  6.24e+02 &  4.67e-24 &  -1.31e+01 &   b\\[0.1ex]
   68 & 1374.1000 &  0.1351 &  6.67e+02 &  4.17e-24 &  -1.55e+01 &  1374.1000 &  0.1330 &  1.54e+06 &  6.03e-23 &  -3.28e+04 &   b\\[0.1ex]
   69 & 1375.8167 &  0.2496 &  8.66e+06 &  7.85e-23 &  -1.60e+05 &  1375.8153 &  0.2509 &  9.71e+02 &  6.76e-24 &  -1.40e+01 &   b\\[0.1ex]
   70 & 1397.9347 &  0.1343 &  1.61e+03 &  5.09e-24 &  -4.25e+01 &  1397.9333 &  0.1356 &  6.32e+02 &  4.83e-24 &  -1.43e+01 &   b\\[0.1ex]
   71 & 1489.2181 &  0.2425 &  1.13e+03 &  4.95e-24 &  -1.80e+01 &  1489.2194 &  0.2418 &  5.96e+02 &  4.96e-24 &  -8.41e+00 &   b\\[0.1ex]
   72 & 1495.2931 &  0.2541 &  1.63e+04 &  9.75e-24 &  -3.03e+02 &  1495.2931 &  0.2568 &  9.95e+03 &  1.02e-23 &  -1.79e+02 &   b\\[0.1ex]
   73 & 1506.0250 &  0.1746 &  7.25e+02 &  4.44e-24 &  -1.41e+01 &  1506.0264 &  0.1767 &  5.38e+04 &  1.56e-23 &  -1.27e+03 &   b\\[0.1ex]
   74 & 1514.1667 &  0.2749 &  2.56e+04 &  1.09e-23 &  -4.50e+02 &  1514.1694 &  0.2756 &  7.15e+02 &  5.28e-24 &  -9.45e+00 &   b\\[0.1ex]
   75 & 1563.0278 &  0.2769 &  9.49e+02 &  5.04e-24 &  -1.26e+01 &  1563.0278 &  0.2796 &  6.74e+02 &  5.86e-24 &  -7.98e+00 &   b\\[0.1ex]
   76 & 1574.2778 &  0.2212 &  5.89e+02 &  4.50e-24 &  -8.53e+00 &  1574.2778 &  0.2226 &  1.31e+03 &  7.40e-24 &  -2.36e+01 &   b\\[0.1ex]
   77 & 1578.4389 &  0.2533 &  5.78e+02 &  4.50e-24 &  -7.81e+00 &  1578.4375 &  0.2561 &  4.26e+03 &  9.95e-24 &  -6.90e+01 &   b\\[0.1ex]
   78 & 1600.3236 &  0.1400 &  1.84e+03 &  6.02e-24 &  -4.90e+01 &  1600.3236 &  0.1428 &  4.13e+02 &  5.26e-24 &  -8.25e+00 &   b\\[0.1ex]
   79 & 1607.9000 &  0.1503 &  1.08e+04 &  9.48e-24 &  -2.90e+02 &  1607.9000 &  0.1476 &  4.63e+02 &  5.38e-24 &  -9.56e+00 &   b\\[0.1ex]
   80 & 1611.5097 &  0.1596 &  4.55e+02 &  4.27e-24 &  -8.86e+00 &  1611.5111 &  0.1582 &  5.75e+02 &  5.59e-24 &  -1.17e+01 &   b\\[0.1ex]
   81 & 1627.7389 &  0.2915 &  6.69e+02 &  4.75e-24 &  -7.99e+00 &  1627.7389 &  0.2887 &  8.37e+02 &  6.36e-24 &  -1.08e+01 &   a\\[0.1ex]
   82 & 1719.2667 &  0.2443 &  1.32e+06 &  4.43e-23 &  -2.68e+04 &  1719.2653 &  0.2429 &  5.50e+05 &  3.61e-23 &  -1.04e+04 &   b\\[0.1ex]
   83 & 1738.7639 &  0.1501 &  2.27e+06 &  4.87e-23 &  -5.90e+04 &  1738.7653 &  0.1473 &  4.15e+02 &  5.47e-24 &  -7.79e+00 &   b\\[0.1ex]
   84 & 1824.0167 &  0.1996 &  8.76e+08 &  1.88e-22 &  -1.89e+07 &  1824.0194 &  0.2016 &  5.26e+02 &  6.27e-24 &  -7.91e+00 &   b\\[0.1ex]
   85 & 1842.9083 &  0.1588 &  3.66e+03 &  8.23e-24 &  -9.06e+01 &  1842.9097 &  0.1615 &  4.50e+02 &  6.17e-24 &  -8.02e+00 &   b\\[0.1ex]
   86 & 1920.0792 &  0.1825 &  4.63e+02 &  5.14e-24 &  -7.97e+00 &  1920.0806 &  0.1853 &  4.92e+02 &  6.58e-24 &  -7.76e+00 &   b\\[0.1ex]
   87 & 1940.8569 &  0.1655 &  5.97e+02 &  5.51e-24 &  -1.22e+01 &  1940.8583 &  0.1683 &  4.59e+02 &  6.19e-24 &  -7.85e+00 &   b\\[0.1ex]
   88 & 1976.2778 &  0.2913 &  2.81e+04 &  1.47e-23 &  -4.64e+02 &  1976.2792 &  0.2886 &  6.62e+02 &  7.06e-24 &  -8.03e+00 &   a\\[0.1ex]
   89 & 2005.5792 &  0.2230 &  6.26e+02 &  5.88e-24 &  -9.18e+00 &  2005.5778 &  0.2216 &  1.46e+04 &  1.54e-23 &  -2.75e+02 &   b\\[0.1ex]
   90 & 2007.7694 &  0.3526 &  2.77e+03 &  8.57e-24 &  -3.74e+01 &  2007.7694 &  0.3526 &  8.51e+02 &  7.55e-24 &  -9.18e+00 &   b\\[0.1ex]
\hline\\[0.1ex]
\caption{All clustered outliers from the Sco X-1 search. Follow-up codes: \textit{(a)}, broad disturbance found in amplitude spectral density (ASD) \textit{(b)}, sharp lines found in ASD; \textit{(c)}, corresponds to known injected signal; \textit{(d)}, followed-up with coherent-summing in Table~\ref{ScoX1S6outlierTable}.\label{AllOutlierTable}}
\end{longtable*}

\bibliography{bibliography.bib}

\end{document}